\documentclass[twocolumn]{aastex61}

\usepackage{amsmath}
\usepackage{graphicx}
\usepackage{float}
\usepackage{epstopdf}

\newcommand{\vek}{\mathbf}

\received{XXX}
\revised{YYY}
\accepted{ZZZ}
\submitjournal{ApJ}
\shorttitle{Implications of GR Centenary Flare in OJ~287}
\shortauthors{Dey et al.}
\begin{document}

%\title{The AAS Journal's Overleaf template using \aastex\ v6.1\footnote{This version fixes many bugs from v6.0 and introduces some new features, primarily in the way the author and affiliations are now marked up.}}
%\title[GR Centenary Flare in OJ~287]{Constraining higher order gravitational wave back reaction in OJ~287 black hole binary using GR centenary flare: improved orbital parameters}
\title[Implications of GR Centenary Flare in OJ~287]{ Authenticating the Presence of a Relativistic Massive Black Hole Binary in OJ~287 Using its General Relativity Centenary Flare: Improved Orbital Parameters}

\correspondingauthor{Lankeswar Dey}
\email{lankeswar.dey@tifr.res.in}

%\author[0000-0002-0786-7307]{Lankeswar Dey}
\author{Lankeswar Dey}
\affiliation{Department of Astronomy and Astrophysics, Tata Institute of Fundamental Research, Mumbai 400005, India\\}

\author{M.~J.~Valtonen}
\affiliation{Finnish Centre for Astronomy with ESO, University of Turku, Finland\\}
\affiliation{Tuorla Observatory, Department of Physics and Astronomy, University of Turku, Finland\\}
%\collaboration{(AAS Journal\newcommand{\vek}[1]{\boldsymbol{#1}} s Data Scientists collaboration)}

\author{A.~Gopakumar}
\affiliation{Department of Astronomy and Astrophysics, Tata Institute of Fundamental Research, Mumbai 400005, India\\}
%\affiliation{AAS Journals Associate Editor-in-Chief}
%\nocollaboration

\author{S.~Zola}
\affiliation{Astronomical Observatory, Jagiellonian University, ul. Orla 171, Cracow PL-30-244, Poland\\}
\affiliation{Mt. Suhora Astronomical Observatory, Pedagogical University, ul. Podchorazych 2, PL30-084 Cracow, Poland\\}
%\collaboration{(LaTeX collaboration)}

\author{R.~Hudec}
\affiliation{Czech Technical University in Prague, Faculty of Electrical Engineering, Technicka 2, Prague 166 27, Czech Republic\\}
\affiliation{Engelhardt Astronomical observatory, Kazan Federal University, Kremlyovskaya street 18, 420008 Kazan, Russian Federation\\}

\author{P.~Pihajoki}
\affiliation{Department of Physics, University of Helsinki, Gustaf H\"allstr\"omin katu 2a, FI-00560, Helsinki, Finland\\}

\author{S.~Ciprini}
\affiliation{Space Science Data Center - Agenzia Spaziale Italiana, via del Politecnico, snc, I-00133, Roma, Italy\\}
\affiliation{Instituto Nazionale di Fisica Nucleare, Sezione di Perugia, Perugia I-06123, Italy\\}

\author{K.~Matsumoto}
\affiliation{Astronomical Institute, Osaka Kyoiku University, 4-698 Asahigaoka, Kashiwara, Osaka 582-8582, Japan\\}

\author{K.~Sadakane}
\affiliation{Astronomical Institute, Osaka Kyoiku University, 4-698 Asahigaoka, Kashiwara, Osaka 582-8582, Japan\\}

\author{M.~Kidger}
\affiliation{Herschel Science Centre, ESAC, European Space Agency, 28691 Villanueva de la Ca{\~n}ada, Madrid, Spain\\}

% ...

%% Note that the \and command from previous versions of AASTeX is now
%% depreciated in this version as it is no longer necessary. AASTeX 
%% automatically takes care of all commas and "and"s between authors names.

%% AASTeX 6.1 has the new \collaboration and \nocollaboration commands to
%% provide the collaboration status of a group of authors. These commands 
%% can be used either before or after the list of corresponding authors. The
%% argument for \collaboration is the collaboration identifier. Authors are
%% encouraged to surround collaboration identifiers with ()s. The 
%% \nocollaboration command takes no argument and exists to indicate that
%% the nearby authors are not part of surrounding collaborations.

%% Mark off the abstract in the ``abstract'' environment. 
\begin{abstract}

Results from regular monitoring of relativistic compact binaries like PSR 1913+16 are consistent with the dominant (quadrupole) order emission of gravitational waves (GWs). We show that observations associated with the binary black hole central engine of blazar OJ~287 demand the inclusion of gravitational radiation reaction effects beyond the quadrupolar order. It turns out that even the effects of certain hereditary contributions to GW emission are required to predict impact flare timings of OJ~287. We develop an approach that incorporates this effect into the binary black hole model for OJ~287. This allows us to demonstrate an excellent agreement between the observed impact flare timings and those predicted from ten orbital cycles of the binary black hole central engine model. The deduced rate of orbital period decay is nine orders of magnitude higher than the observed rate in PSR 1913+16, demonstrating again the relativistic nature of OJ~287's central engine. Finally, we argue that precise timing of the predicted 2019 impact flare should allow a test of the celebrated black hole ``no-hair theorem" at the $10\%$ level.

\end{abstract}

%% Keywords should appear after the \end{abstract} command. 
%% See the online documentation for the full list of available subject
%% keywords and the rules for their use.

%% From the front matter, we move on to the body of the paper.
%% Sections are demarcated by \section and \subsection, respectively.
%% Observe the use of the LaTeX \label
%% command after the \subsection to give a symbolic KEY to the
%% subsection for cross-referencing in a \ref command.
%% You can use LaTeX's \ref and \label commands to keep track of
%% cross-references to sections, equations, tables, and figures.
%% That way, if you change the order of any elements, LaTeX will
%% automatically renumber them.

%% We recommend that authors also use the natbib \citep
%% and \citet commands to identify citations.  The citations are
%% tied to the reference list via symbolic KEYs. The KEY corresponds
%% to the KEY in the \bibitem in the reference list below. 

\section{Introduction} \label{sec:intro}

\begin{figure}
\centering
\includegraphics[width=\linewidth]{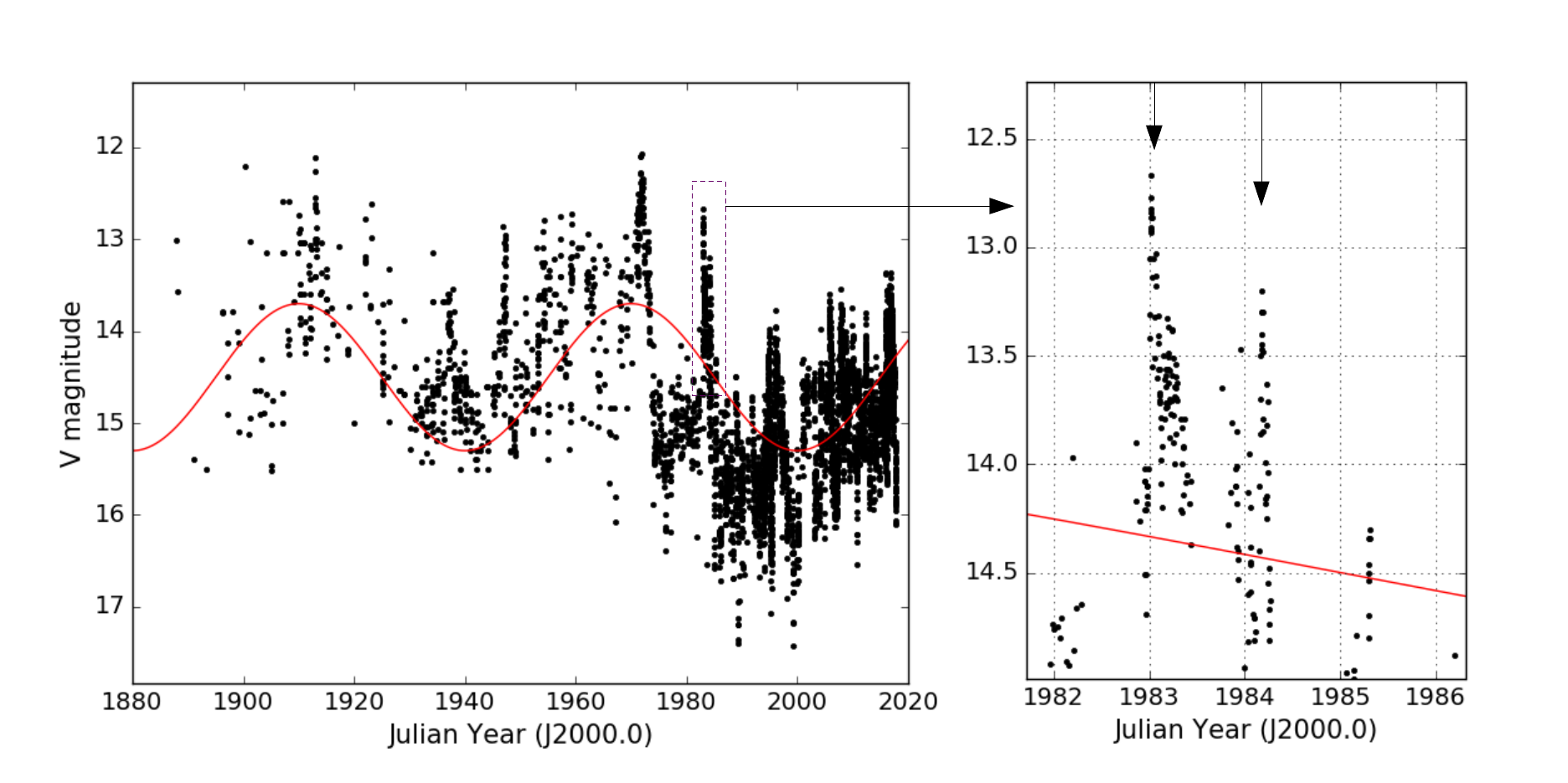}
\caption{ The left panel displays the optical light curve of OJ~287 from 1886 to 2017. We draw a fiducial curve for easy visualization of the inherent long-term variations. The right panel shows the observed double-peaked structure of the high-brightness flares. The positions of the two peaks are indicated by downward arrows from the top of the panel.}
\label{fig:lightcurve}
\end{figure}

OJ~287 (RA: 08:54:48.87  \& DEC:+20:06:30.6) is a bright blazar, a class of active galactic nuclei, situated near the ecliptic in the constellation of Cancer. This part of the sky has been frequently photographed for other purposes since late 1800's and therefore it has been possible to construct an exceptionally long and detailed light curve for this blazar using the historical plate material. 
It is at a redshift ($z$) of 0.306 corresponding to a luminosity distance of 1.6 Gpc in the standard $\Lambda$CDM cosmology which makes it a relatively nearby object as blazars go.  
The optical light curve, extending over 120 yr \citep{sil88,hud13}, exhibits repeated high-brightness flares (see Figure~\ref{fig:lightcurve}). A visual inspection reveals the presence of two periodic variations with approximate timescales of 12 yr and 60 yr which  have been confirmed through a quantitative analysis \citep{val06}.
We mark the  $\sim 60$ year periodicity by a red curve in the left panel of Figure~\ref{fig:lightcurve} and many observed outbursts/flares are separated by $\sim 12$ years.
The regular monitoring of OJ~287, pursued only in the recent past, reveal that these outbursts come in pairs and are separated by a few years.
The doubly peaked structure is shown in the right panel of Figure~\ref{fig:lightcurve}. 
The presence of double periodicity in the optical light curve provided an early evidence for the occurrence of quasi-Keplerian orbital motion in the blazar, where the 12 year periodicity corresponds to the orbital period timescale and the 60 year timescale is related to the orbital precession. The ratio of the two deduced periods gave an early estimate for the total mass of the system to be $\sim 18\times 10^9\, M_{\odot}$, provided we invoke general relativity to explain the orbital precession  \citep{pie98}.
It is important to note that this estimate is quite independent of the detailed central engine properties of OJ~287. The host galaxy is hard to detect because of the bright nucleus; however, during the recent fading of the nucleus by more than two magnitudes below the high level state it has been possible to get a reliable magnitude of the host galaxy. It turns out to be similar to NGC~4889 in the Coma cluster of galaxies, i.e. among the brightest in the universe. These results will be reported elsewhere (Valtonen et al. 2018).
These considerations eventually led to the development of the binary black hole (BBH) central engine model for OJ~287 \citep{leh96,val08a}.  

\begin{figure}
\centering
\includegraphics[width=\linewidth]{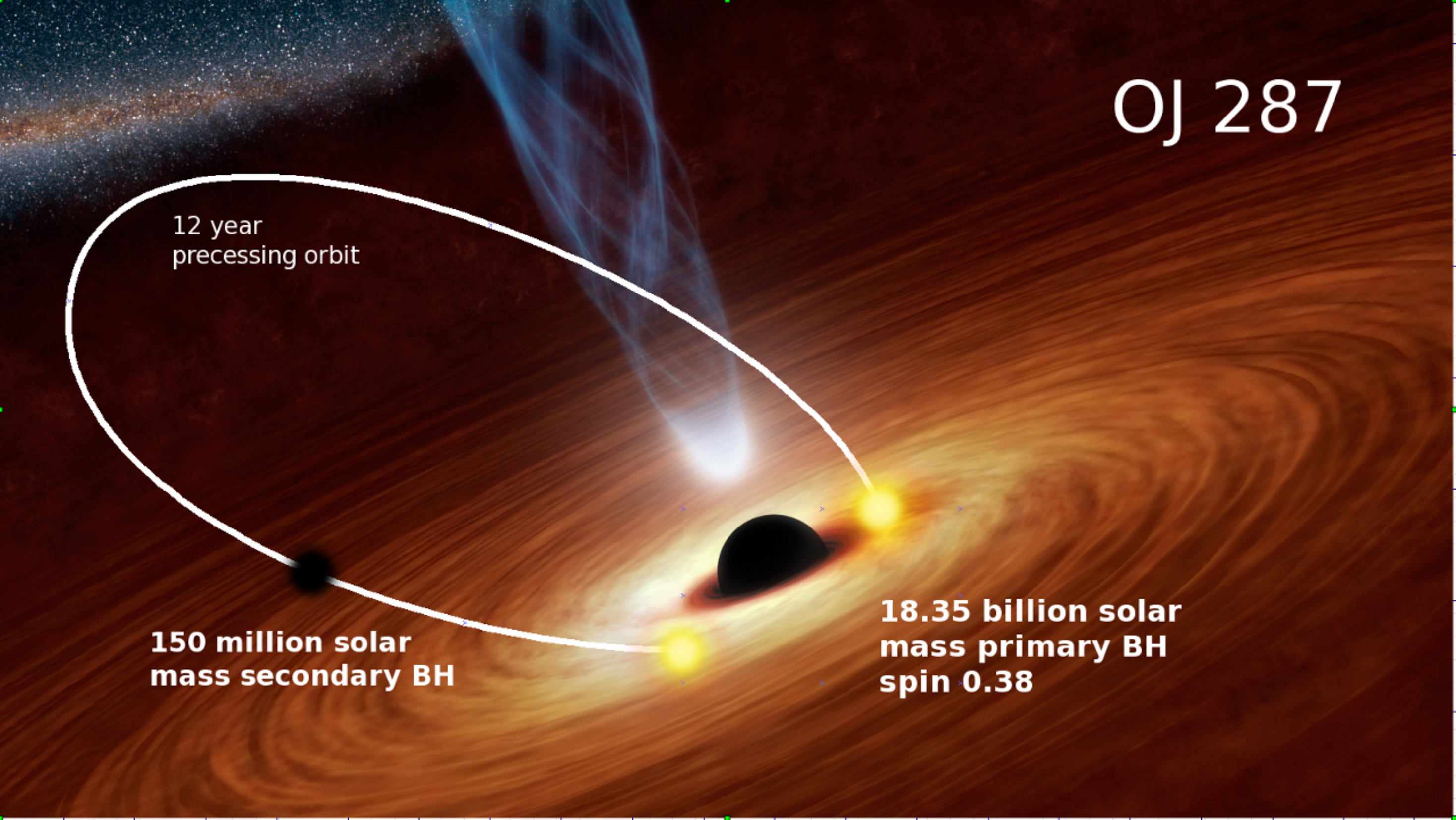}
\caption{ Artistic illustration of the binary black hole system in OJ~287.
The present analysis provides an improved estimate for the spin of the primary black hole.
\label{fig:model}}
\end{figure}

According to the BBH model, the central engine of OJ~287 contains a binary black hole system where a super-massive secondary black hole is orbiting an ultra-massive primary black hole in a  precessing eccentric orbit with a redshifted orbital period of $\sim 12$ yr (see Figure~\ref{fig:model}). 
The primary cause of certain observed flares (also called outbursts) in this model is the impact of the secondary black hole on the accretion disk of the primary \citep{leh96,pih16}. The impact forces the release of two hot bubbles of gas on both sides of the accretion disk which radiate strongly after becoming optically transparent, leading to a sharp rise in the apparent brightness of OJ~287. 
The less massive secondary BH impacts the accretion disk twice every orbit while traveling along a precessing eccentric orbit (Figure~\ref{fig:model}). This results in double-peaked quasi-periodic high-brightness (thermal) flares from OJ~287.  
Furthermore, large amounts of matter get ejected from the accretion disk during the impact and are subsequently accreted to the disk center. This ensures that part of the unbound accretion-disk material ends up in the twin jets. %Reference?
The matter accretion leads to non-thermal flares via relativistic shocks in the jets which produce the secondary flares in OJ~287, lasting more than a year after the first thermal flare \citep{val09}.

The BBH model of OJ~287 can be used to predict the flare timings \citep{sun97,val08b,val11b} and the latest prediction was successfully verified in 2015 November. The optical brightness of OJ~287 rose above the levels of its normal variations on November 25, and it achieved peak brightness on December 5. On that date, OJ~287 was brighter than at any time since 1984 \protect\citep{val16}. Owing to the coincidence of the start of the flare with the date of completion of general relativity (GR) by Albert Einstein one hundred years earlier, it was termed as the GR centenary flare. Detailed monitoring of the 2015 impact flare allowed us to  estimate the spin of the primary BH to be $\sim 1/3$ of the maximum value allowed in GR. 
This was the fourth instance when multi-wavelength observational campaigns were launched to observed predicted impact flares from the BBH central engine of OJ~287 \citep{val08b,val16}. The latest observational campaigns confirmed the presence of a spinning massive BH binary inspiraling due to the emission of nano-Hertz gravitational waves in OJ~287. 
These developments influenced the Event Horizon Telescope consortium to launch observational campaigns in 2017 and 2018 to resolve the presence of two BHs in OJ~287 via the millimeter wavelength Very Long Baseline Interferometry.

Predictions of impact flare timings are made by solving post-Newtonian (PN) equations of motion to determine the  secondary BH orbit around the primary while using the observed outburst times as fixed points of the orbit. 
The PN approximation  provides general relativistic corrections to Newtonian dynamics in powers of $(v/c)^2$, where $v$ and $c$ are the characteristic orbital velocity and the speed of light, respectively. The GR centenary flare was predicted using 3PN-accurate (i.e., third PN order) BBH dynamics that employed GR corrections to Newtonian dynamics accurate to order $(v/c)^6$ \citep{val10a,val10b,val11a}. Additionally, earlier investigations invoked nine fixed points in the BBH orbit, which allowed the unique determination of eight parameters of the OJ~287 BBH central engine model \citep{val10a, val11a}. The GR centenary flare provided the tenth fixed point of BBH orbit, which opens up the possibility of constraining an additional parameter of the central engine.
Moreover, the GW emission-induced rate of orbital period decay of the BBH in OJ~287, estimated to be $\sim 10^{-3}$, makes it an interesting candidate for probing the radiative sector of relativistic gravity \citep{Wex14}.

These considerations influenced us to explore the observational consequences of incorporating even higher-order PN contributions to the BBH dynamics. Therefore, we introduce the effects of GW emission beyond the quadrupolar order on the dynamics of the BBH in OJ~287 while additionally incorporating next-to-leading-order spin effects \citep{LB_review, BBF_06,wil17}.
Moreover, we incorporate the effects of dominant order hereditary contributions to GW emission, detailed in \cite{BS93}, on to the binary BH orbital dynamics.
It turns out that these improvements to BBH orbital dynamics cause non-negligible changes to our earlier estimates for the BBH parameters, especially for the dimensionless angular momentum parameter of the primary BH in OJ~287, and the inclusion of present improvements to BBH orbital dynamics should allow the test of the black hole ``no-hair theorem" during the present decade. This is essentially due to our current  ability to accurately  predict the time of the next impact flare from OJ~287, influenced by the present investigation.

This paper is structured as follows. In Section~\ref{sec:pnd}, we discuss briefly the improved BBH orbital dynamics. Details of  our approach to obtain the parameters of the  BBH system from optical observation of OJ~287 are presented in Section~\ref{sec:bbhorbit}. How we incorporate the effects of dominant-order ``hereditary" contributions to GW emission into  BBH dynamics is detailed in Section~\ref{sec:radfac}.
Implications of our improved BBH model on historic and future observations are outlined in Section~\ref{sec:lc_comparison}. In Appendix~\ref{app:kdndldetdl}, we display PN-accurate expressions used to incorporate ``hereditary" contributions to BBH dynamics.

\section{Post-Newtonian-Accurate BBH Dynamics} \label{sec:pnd}

The PN approach, as noted earlier, provides general relativistic corrections to Newtonian dynamics in powers of $(v/c)^2$. In this paper, we deploy a PN-accurate expression for the relative acceleration in the center-of-mass frame, appropriate for compact binaries of arbitrary masses and spins. Influenced by \citet{LB_review,wil17}, we schematically write 
\begin{eqnarray} 
\ddot{\vek x} \equiv \frac{d^2{\vek x}} {dt^2} &=& \ddot{{\vek x}}_{\rm 0} + \ddot{{\vek x}}_{\rm 1PN} + \ddot{{\vek x}}_{\rm 2PN} + \ddot{{\vek x}}_{\rm 3PN} \nonumber\\ 
&& + \ddot{{\vek x}}_{\rm 2.5PN} + \ddot{{\vek x}}_{\rm 3.5PN} + \ddot{{\vek x}}_{\rm 4PN(tail)} + \ddot{{\vek x}}_{\rm 4.5PN} \nonumber \\
&& + \ddot{{\vek x}}_{\rm SO} + \ddot{{\vek x}}_{\rm SS}  + \ddot{{\vek x}}_{\rm Q} + \ddot{{\vek x}}_{\rm 4PN(SO-RR)}, 
\label{eqn:eom}
\end{eqnarray} 
where ${\vek x} = {\vek x}_1 - {\vek x}_2$ gives the center-of-mass relative separation vector between the black holes with masses $m_1$ and $m_2$. The familiar Newtonian contribution, denoted by $\ddot{{\vek x}}_{0}$, is given by $\ddot{{\vek x}}_{0} = -\frac{ G\, m}{r^3} \, {\vek x}$, where $m= m_1 + m_2$, $r = |{\vek x}|$. Additionally, below we use $\hat{\vek n} \equiv {\vek x}/r$, $\dot{{\vek x}} = {\vek v}$ and $\eta = m_1\, m_2/m^2$.
The PN contributions occurring at 1PN, 2PN, and 3PN orders, denoted by $\ddot{{\vek x}}_{\rm 1PN}$, $\ddot { {\vek x}}_{\rm 2PN}$, $\ddot { {\vek x}}_{\rm 3PN}$, are conservative in nature and result in a precessing eccentric orbit. The explicit expressions for these contributions can easily be adapted from Equations~(219)-(222) in \cite{LB_review} and therefore are in the {\it modified harmonic gauge}.
The second line contributions enter the $ d {\vek x}/dt$ expression at 2.5PN, 3.5PN, 4PN, and 4.5PN orders and are respectively denoted by $\ddot{{\vek x}}_{\rm 2.5PN}$, $\ddot{{\vek x}}_{\rm 3.5PN}$, $\ddot{{\vek x}}_{\rm 4PN(tail)}$, and $\ddot{{\vek x}}_{\rm 4.5PN}$. These are reactive terms in the orbital dynamics and cause the shrinking of BBH orbit due to the emission of GWs, and their explicit expressions are available in Equations~(219) and (220) of \cite{LB_review}. Later, we will provide explanations for the $\ddot{{\vek x}}_{\rm 4PN(tail)}$ term in detail.

The conservative spin contributions enter the equations of motion via spin-orbit and spin-spin couplings and are listed in the third line of Equation~(\ref{eqn:eom}). These are denoted by $\ddot{\vek x}_{\rm SO}$ and $\ddot{\vek x}_{\rm SS}$, while the $\ddot{{\vek x}}_{\rm Q}$ term stands for a classical spin-orbit coupling that brings in the quadrupole deformation of a rotating BH. The term $ \ddot{{\vek x}}_{\rm 4PN(SO-RR)}$ stands for the spin-orbit contribution to the gravitational radiation reaction, extractable from Equation~(8) in \cite{ZW07}.
%http://adsabs.harvard.edu/abs/2007GReGr..39.1661Z
We adapted Equations~(5.7a) and (5.7b) of \cite{BBF_06} to incorporate spin-orbit contributions that enter the dynamics at 1.5PN and 2.5PN orders, and these equations generalize the classic result of \cite{bar75}. The dominant-order general relativistic spin-spin and classic spin-orbit contributions, entering the $\ddot{{\vek x}}$ expression at 2PN order, are extracted from \cite{val10a}, and we have verified that our explicit expressions are consistent with Equation~(2.3) of \cite{wil17}.

The spin of the primary black hole precesses owing to general relativistic spin-orbit, spin-spin, and classical spin-orbit couplings, and the relevant equation for ${\vek s}_1$, the unit vector along the direction of primary BH spin, may be symbolically written as
\begin{subequations}
\begin{eqnarray}
\frac{d{\vek s}_1}{dt} &=& {\vek \Omega} \times {\vek s}_1 \,, \\
{\vek \Omega} &=& {\vek \Omega}_{\rm SO} + {\vek \Omega}_{\rm SS} + {\vek \Omega}_{\rm Q} \,,
\end{eqnarray}
\label{eqn:spinprecession}
\end{subequations}
\noindent
where the spin of the primary black hole in terms of its Kerr parameter ($\chi_1$) is given by ${\vek S}_1 = G\, m_1^2 \, \chi_1 \, {\vek s}_1/c$ ($\chi_1$ is allowed to take values between $0$ and $1$ in GR). For the general relativistic spin-orbit contributions to ${\vek \Omega}$, we have adapted Equations~(6.2) and (6.3) of \cite{BBF_06}, while spin-spin and classical spin-orbit contributions are listed by \cite{val11a}.

Let us turn our attention on the radiation reaction (RR) terms, listed in the second line of Equation~(\ref{eqn:eom}). The radiation reaction contributions to $\ddot{\vek x}$ appearing at 2.5PN, 3.5PN, and 4.5PN orders can be written as
\begin{eqnarray}
\ddot{{\vek x}}_{\rm iPN} =& -\frac{8}{5} \frac{G^2 m^2 \eta}{c^{2i}\, r^3} \left(A_{\rm i}\, \dot{r} {\vek n} - B_{\rm i}\, {\vek v}\right) \label{eqn:rr_2p5-4p5}
\end{eqnarray}
where i can take the values 2.5, 3.5 and 4.5. The A and B coefficients in Equation~(\ref{eqn:rr_2p5-4p5}) are calculated by employing the balance argument of \cite{IW95} that equate appropriate PN-accurate time derivatives of ``near-zone" orbital energy and angular momentum  expressions to  PN-accurate ``far-zone" GW energy and angular momentum fluxes. This balance approach of \cite{IW95} introduces some independent degrees of freedom in the RR terms (2 degrees of freedom for 2.5PN, 6 for 3.5PN and 12 for 4.5PN) and we use {\it harmonic gauge} for fixing these independent parameters. The dominant 2.5PN order contributions in {\it harmonic gauge}, available in \cite{IW95}, reads
\begin{subequations}
\begin{eqnarray} 
A_{2.5} =&  3 v^2 +  \frac{17}{3} \frac{G\,m}{r} \\
B_{2.5} =&   v^2 + 3 \frac{G\,m}{r}\, .
\end{eqnarray} 
\end{subequations}
\noindent
The explicit  expressions for the 3.5PN order contributions in {\it harmonic gauge} can be extracted from Equations~(219) and (220) in \cite{LB_review}, and we invoked \cite{gop97} for the 4.5PN order contributions. 
It turns out that these A-coefficients do not contribute to the secular evolution of binary BH orbit. This is mainly because they are nearly symmetric but opposite in sign with respect to the pericenter while integrating over a quasi-Keplerian orbit.
In contrast, the B-coefficients do not suffer such sign changes with respect to the pericenter and therefore contribute to the the secular BBH orbital evolution.

A few comments on the {\it balance arguments} of \cite{IW95} are in order. The method crucially requires explicit closed-form expressions for the ``far-zone" GW energy and angular momentum fluxes, valid for non-circular orbits. This is why the fully 2PN-accurate ``instantaneous" contributions to GW energy and angular momentum fluxes, derived by \cite{GI97}, provided radiation reaction contributions $\ddot{\vek x}$ at 2.5PN, 3.5PN, and 4.5PN orders \citep{gop97}.
The ``instantaneous" labeling is influenced by \cite{BDI} that recommended the split of higher-PN-order far-zone fluxes into two parts.
The contributions that purely depend on the state of the binary at the retarded instant are termed as the ``instantaneous" contributions while those contributions that are {\it a priori} sensitive to the whole past orbits of the binary are called ``tails" or ``hereditary" contributions. 
These tail contributions are usually expressed in terms of integrals extending over the whole  past ``history" of the binary and therefore it is not possible to find closed-form expressions for far-zone energy and angular momentum fluxes as demonstrated by \cite{BS93,RS97}. 

Incidentally, the dominant-order tail contributions to far-zone fluxes are $(v/c)^3$ corrections to the quadrupolar order GW fluxes which can potentially contribute to $(v/c)^8$ terms in the orbital dynamics. 
Unfortunately, it is not possible to compute such reactive contributions using the above-mentioned {\it balance arguments} of \cite{IW95} and there exist no explicit closed-form expressions for $\ddot{{\vek x}}_{\rm 4PN(tail)}$ for compact binaries in non-circular orbits.
This is essentially because of the nonavailability of closed-form expressions for the dominant-order tail contributions to energy and angular momentum fluxes as noted earlier.

This forced us to introduce a heuristic way of incorporating the effect of dominant-order tail contributions to GW emission into  BBH orbital dynamics. We implement it by introducing an ambiguity parameter $\gamma$ at the dominant-order radiation reaction terms such that the second line of Equation~(\ref{eqn:eom}) becomes
\begin{eqnarray}
\ddot{{\vek x}}_{\rm 2.5PN} + \ddot{{\vek x}}_{\rm 3.5PN}  + \ddot{{\vek x}}_{\rm 4PN(tail)} + \ddot{{\vek x}}_{\rm 4.5PN} = \gamma\, \ddot{{\vek x}}_{\rm 2.5PN} \nonumber \\ + \ddot{{\vek x}}_{\rm 3.5PN}  + \ddot{{\vek x}}_{\rm 4.5PN}\,.
\label{eqn:gamma_tail}
\end{eqnarray}
Clearly, the value of $\gamma$ will have to be determined from outburst observations of OJ~287.
In Section~\ref{sec:radfac}, we demonstrate that the observationally determined $\gamma$ value ($\gamma_{obs}$) is fully consistent with the general relativistic orbital phase evolution of the BBH present in OJ~287. This is achieved by adapting certain GW phasing formalism, developed for constructing PN-accurate inspiral GW templates for comparable-mass compact binaries \citep{DGI,KG06}. 
The physical reason for incorporating the effect of dominant order `tail' contributions to $\ddot{{\vek x}}$ in an heuristic way will be discussed in subsection~\ref{subsec:tail_term_importance}.

The fact that we are able to fix an appropriate general relativistic value for the  ambiguity parameter $\gamma$ (denoted by $\gamma_{GR}$) prompted us to explore the possibility  of testing the celebrated black hole no-hair theorem \citep{Hansen74}.
We are influenced by the direct consequence of the BH no-hair theorem which demands that the dimensionless quadrupole moment (${q_2}$) of a general relativistic BH should be related to its Kerr parameter ($\chi$) by a simple relation $ {q_2} = -\, \chi^2$ \citep{Thorne80}.
%http://adsabs.harvard.edu/abs/1980RvMP...52..299T
This idea is implemented by introducing an additional parameter $q$ to characterize the classical spin-orbit contributions to the BBH equations of motion \citep{bar75} such that 
\begin{eqnarray}
\ddot{{\vek x}}_{\rm Qnew} = q\ {\chi}^2 \ \frac{\displaystyle 3\,G^3\,{m_1}^2\,m}{\displaystyle 2\,c^4\,r^4} \Bigl\{ \Bigl[ 5 ({\vek n} \,.\, {\vek s_1})^2 -1 \Bigr] {\vek n} \nonumber \\ - 2 ({\vek n} \,.\, {\vek s_1}) {\vek s_1}\Bigr\},
\end{eqnarray}
where we have replaced the scaled quadrupole moment by $-q\,\chi^2$, and in GR the value of $q$ should be unity. 
The proposed test involves determining the value of $q$ from the accurate timing of the next impact flare, expected to peak on 2019 July 31.

The present effort neglected the frictional energy loss due to the passage of secondary BH through the accretion disk of the primary BH. This is justified as the frictional energy loss is much smaller than its GW counterpart. To clarify the claim, we note that $\sim 16\, M_{\odot}$ of matter is extracted from the accretion disk due to the passage of the secondary BH \citep{pih13}. This forces a change in the momentum of the secondary and the fractional momentum loss ($\Delta p_s/p_s$) is $\sim 10^{-7}$ per encounter, or $\sim 2 \times 10^{-7}$ per orbit. The associated frictional energy loss is $\sim 4 \times 10^{-7}$ per orbit and it leads to a rate of orbital period change $\dot{P_b} \sim 6 \times 10^{-7}$. In contrast, GW emission induced rate of orbital period change is  $\sim 10^{-3}$. This shows that the effect of GW emission is four orders of magnitude higher than its frictional counterpart which is not surprising as the secondary BH spends very little time ($\sim 3\%$ of its orbital period) crossing the accretion disk whereas the energy loss due to GW builds up during the whole orbit.
In the next section, we explain in detail our approach to determine the BBH central engine parameters from the observed impact flare timings.

\section{ Determining the Relativistic BBH Orbit of OJ~287} \label{sec:bbhorbit}

This section details our approach to  determine the  parameters of OJ~287's BBH central engine, depicted in Figure~\ref{fig:model}. We use the  {\it accurately} extracted (observed) starting epochs of ten optical outbursts of OJ~287 {to track the binary orbit}. In the BBH model, these epochs correspond to ten ``fixed points" of the orbit that lead to nine time intervals. We use these nine intervals to determine nine independent parameters that describe the BBH central engine of OJ~287. The adopted outburst timings with uncertainties are displayed in Table~\ref{tab:outburst}, while the relevant sections of the observed light curve at these epochs are shown in Section~\ref{sec:lc_comparison}.

\startlongtable
\begin{deluxetable}{c}
\tablecaption{ Extracted starting times (in Julian year) of the observed optical outbursts of OJ~287.
The data points prior to 1970 are extracted from archival photographic plates while the historical 1913 flare time is according to \cite{hud13}.
\label{tab:outburst}  }
\tablehead{
\colhead{Outburst times with estimated uncertainty}}
\startdata
1912.980    $\pm$  0.020\\ 
1947.283    $\pm$  0.002\\ 
1957.095    $\pm$  0.025\\ 
1972.935    $\pm$  0.012\\
1982.964    $\pm$  0.0005\\  
1984.125    $\pm$  0.01\\ 
1995.841    $\pm$  0.002\\ 
2005.745    $\pm$  0.015\\ 
2007.6915   $\pm$  0.0015\\ 
2015.875    $\pm$  0.025\\ 
\enddata
\end{deluxetable}

\subsection{ Model for OJ~287's Central Engine and its 
Implementation} \label{subsec:orbit solving}

\begin{figure}
\plotone{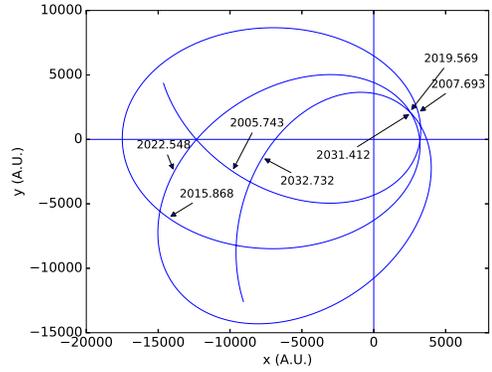}
\caption{Typical orbit of the secondary BH in OJ~287 in the 2005--2033 window.
The primary BH is situated at the origin with its accretion disk in the $y=0$ plane. The locations of the secondary BH at the time of different outburst epochs are marked by arrow symbols. The time-delay effect is clearly visible, while close inspection reveals that these delays for different impacts are different.
The use of Equation~(\ref{eq_time_delay}) ensures that the values for $d$ and $h$ should remain constant if observations are consistent with our  model. 
\label{fig:orbit}}
\end{figure}

Our approach to determine the parameters of BBH  central engine model for OJ~287 proceeds as follows. First, an approximate orbit of the secondary BH is calculated by numerically integrating the above-mentioned  PN-accurate equations of motion (Equation~\ref{eqn:eom}) while using some trial values of the independent parameters. This orbit produces a list of reference times at which the secondary crosses the $y = 0$ plane  of the accretion disk (see Figure~\ref{fig:orbit}).
However, these plane-crossing epochs are not the same as the observed outburst times. We need to take into consideration certain astrophysical processes that occur during the time interval between the BH impact and the observed optical outburst epoch. The effects of these processes are incorporated by adding a ``time delay" to the plane-crossing times.
These delays represent the time interval between the actual creation of a hot bubble of gas due to the  disk impact and when it becomes optically thin and releases a strong burst of optical radiation. An additional temporal correction is required to model the fact that when the secondary black hole approaches the accretion disk, the disk as a whole is pulled toward the secondary. This ensures that the secondary BH impact occurs before it reaches  
the accretion-disk plane  of the primary black hole, depicted by the $y = 0$ line in Figure~\ref{fig:orbit}. 
Therefore, we subtract a time interval, termed ``time advance," from the plane-crossing time. This leads to a new list of corrected reference times.

Let us digress briefly to introduce the way we model the time delay and the accretion disk of the primary BH.
We use the accretion-disk model detailed by \cite{leh96}, which is based on the $\alpha_{g}$ disk model of \cite{sakimoto81}, with scaling provided by \cite{stella84}. In this model, the disk impacts are followed by thermal flares after a time delay $t_d$ given by
\begin{eqnarray}
\label{eq_time_delay}
t_d = d\, {m_2}^{1.24}\, {v_{\rm rel}}^{-4.23}\, h^{-0.29}\, {\Sigma}^{0.91},
\label{eqn:time_delay}
\end{eqnarray}
where the delay parameter $d$ is a proportionality factor to be determined as part of the orbit solution.
This also applies to the disk thickness $h$ and the secondary BH mass $m_2$. The impact velocity of the secondary relative to the disk ($v_{\rm rel}$) is known in the model for each impact and the fiducial values are those given by \citet{leh96}. 
Furthermore, the scaling for the disk surface density $\Sigma$ is
\begin{eqnarray}
\Sigma \approx {\alpha_g}^{-0.8} \, {\dot{m}}^{0.6}, 
\end{eqnarray}
where $\alpha_g$ is the viscosity coefficient and $\dot{m}$ is the mass accretion rate in Eddington units. Typical particle number density in the accretion disk in our model is $\sim 10^{14}\, {\rm cm^{-3}}$ (see Table~2 of \cite{leh96} for detailed astrophysical properties of the disk).
The value of $\alpha_g$ depends on the un-beamed total luminosity of OJ~287: $\alpha_g = 0.1,\, 0.3,\, 1.0$ for the total luminosity of $2.5\times 10^{45}$, $1.2\times 10^{46}$, $5\times 10^{46}$ erg~cm$^{-2}$~s$^{-1}$, respectively. Since the observed (beamed) luminosity is $\sim 10^{47}$ erg~cm$^{-2}$~s$^{-1}$, the most likely $\alpha_g$ value is near the lowest quoted value, $\alpha_g\approx 0.1$, since the relativistic Doppler boosting factor is likely to be in excess of $\delta \approx 20$ \citep{wor82}.
Interestingly, the orbit determination provides a $\alpha_g-\dot{m}$ correlation as a side result while determining the orbit from impact flare timings. However, it is not possible to extract these two parameters individually, since the time delay is practically a function of $\dot{m}/\alpha_g$ and depends weakly on either parameter.

 We now move to work on the above-mentioned corrected reference times. It is customary to normalize the list so that the 1983 outburst has the exact time of 1982.964. This is done by subtracting the difference between the 1983 corrected reference time, namely the ``disk-crossing time plus time delay minus time advance," and the actual 1983 outburst time (1982.964), from all other reference times. 
Thereafter, we check the timing of a certain outburst, typically that of the 1973 outburst, by adjusting usually the initial orientation of the major axis of the binary.
We pursue new trial solutions until the {1973} outburst time is within the observed time interval { ($1972.935 \pm 0.012$)}. 

In the next step, the disk thickness parameter ($h$) is found by requiring that the 2005 outburst timing matches with the observations {($2005.745 \pm 0.015$)}. 
 This process is repeated  until we determine all nine independent parameters of the BBH system. In other words, the procedure involves adjusting  each parameter of the BBH central engine model so that some particular outburst happens within a certain time window. 
Further details of the solving procedure are described by \cite{val07}. At each stage of the iterative procedure, we ensure that the previous conditions are still satisfied; if not, the procedures are repeated. When all the outburst timings, listed in Table~\ref{tab:outburst}, match within the listed uncertainties, we regard that solution as an {\it acceptable solution}.

\subsection{ Extraction of the BBH Central Engine Parameters } \label{subsec:orbit parameters}

We performed 1000 trials for orbit solutions and at each time, as expected,  with a little different initial parameter values.
It turns out that 285 cases converged to an acceptable  solution, but the remaining trials were interrupted as the procedure exceeded the preset number of attempts while varying a parameter. The general experience with the code is that the convergence is not always found even if the trial is continued much longer. 
The average values of the parameters are listed in Table~\ref{tab:parameters}, as well as the $1\sigma$ scatter of these values as the uncertainty. The independent parameters of Table~\ref{tab:parameters} are the two masses $m_1$ and $m_2$, the primary BH Kerr parameter $\chi_1$, the apocenter eccentricity $e_0$, the angle of orientation of the semimajor axis of the orbit $\Theta_0$ in 1856 (the starting year of the orbit calculation), the precession rate of the major axis per period $\Delta\Phi$, and an ambiguity parameter $\gamma_{obs}$ that we employ to incorporate the effects of dominant order hereditary contributions to GW emission in the BBH dynamics, as evident in Equation~(\ref{eqn:gamma_tail})
(we use the subscript $obs$ to distinguish the observationally determined 
$\gamma$ from its GR based estimate).
Additionally, two independent parameters incorporate the effects of astrophysical processes that are associated with the accretion disk impact of the secondary BH. These are listed in Table~\ref{tab:parameters} as $d$ and $h$, where $d$ is the delay parameter present in Equation~(\ref{eqn:time_delay}) while the disk thickness parameter $h$ is a scale factor with respect to the ``standard'' model of \cite{leh96}.
In other words, the average half thickness of the accretion disk is $\sim 3\times 10^{15}$ cm and we need to multiply the disk thickness, given in Table~2 of \cite{leh96}, by disk thickness parameter $h$ to get the actual thickness profile of the disk in our model. 
Note that these two parameters ($d$ and $h$) are functions of the mass accretion rate $\dot m$ and the viscosity parameter $\alpha_g$ of the standard $\alpha_g$ accretion disk.

\startlongtable
\begin{deluxetable}{c|c c c c}
\tablecaption{
Independent and dependent parameters of the BBH system in OJ~287 according to our orbit solution.
Note that $\gamma_{obs}$ provides the observationally 
determined value for the $\gamma$ parameter, invoked to incorporate 
heuristically the effect of dominant order `tail' contributions 
to GW emission on our Equation~(\ref{eqn:gamma_tail}) for BBH dynamics.
%We have also listed some parameters derived from the solution. Note that $\gamma_{obs}$ is the parameter we used in  to model the effects of the {\it tail} term in the dynamics.} 
\label{tab:parameters}}
\tablehead{
\colhead{} & \colhead{Parameter} & \colhead{Value} & \colhead{unit} &\colhead{error}}
\colnumbers
\startdata
 & $m_1$ & 18348 & $10^6\, M_{\odot}$ & $\pm$7.92\\
 & $m_2$ & 150.13 & $10^6\, M_{\odot}$ & $\pm$0.25\\
 & $\chi_1$ & 0.381 & & $\pm$0.004\\
 & $h$ & 0.900 & & $\pm$0.001\\
 Independent& $d$ & 0.776 & & $\pm$0.004\\
 & $\Delta\Phi$ & 38.62 & $deg$ & $\pm$0.01 \\
 & $\Theta_0$ & 55.42  & $deg$ & $\pm$0.17  \\
 & $e_0$ & 0.657 & & $\pm$0.001\\
 & $\gamma_{obs}$ & 1.304 & & $\pm$0.008\\
\hline
Derived & $P^{2017}_{\rm orb}$ & 12.062 & yr & $\pm$0.007\\
 & $\dot{P}_{\rm orb}$ & 0.00099 &  & $\pm$0.00006\\
\enddata
\end{deluxetable}

Table~\ref{tab:parameters} also lists certain derived parameters that characterize the BBH in OJ~287 --- namely, the present (redshifted) orbital period $P^{2017}_{\rm orb} $ and its rate of decrease due to the emission of GWs ($\dot{P}_{\rm orb} $). 
We find the rate of orbital period shrinkage to be $\sim 10^{-3}$; in contrast, the measured $\dot{P}_{\rm orb} $ values for relativistic binary pulsar systems like PSR J1913+16 are $\sim 10^{-12}$ \citep{Wex14}.
%http://adsabs.harvard.edu/cgi-bin/bib_query?arXiv:1402.5594
This is roughly nine orders of magnitude smaller than in OJ~287, which demonstrates the strong-field relativistic nature of  OJ~287. 
To probe the relevance of higher-order radiation reaction terms in Equation~(\ref{eqn:eom}), we repeat the above detailed orbital fitting procedure while employing only the dominant 2.5PN order  contributions to $\ddot{\bf x} $.
This resulted in $\dot{P_{\rm orb} } =0.00106 $, which indicates that the higher-order radiation reaction contributions reduce the quadrupolar-order GW flux by about $6.5\%$.

We demonstrate the predictive power of our BBH central engine model for OJ~287 with the help of  Table~\ref{tab:orbit solution}, which lists all the epochs associated with past impacts as well as future impacts within the years 1886 to 2056 according to our model. 
The entries of Table~\ref{tab:orbit solution} quantify many facets of our model. Column~1 provides the starting times of the outbursts ($t_{\rm out}$) in Julian year (J2000.0), while $t_{\rm del}$ indicates the time delay between the impact of the secondary BH with the accretion disk and the starting of the outburst. The listed $t_{\rm del}$ values differ from those of \cite{leh96} by the scale parameter $d$.
The time advance $(t_{\rm adv})$ arises from the bending of the accretion disk due to the presence of the secondary BH prior to the impact.
The radial distance ($R_{\rm imp}$) of the secondary and its orbital speed ($v_0$) at various impact flare epochs are also listed in Table~\ref{tab:orbit solution}. 
In a later section, we will explain the importance of the next predicted outburst.

\startlongtable
\begin{deluxetable}{c c c c c}
\tablecaption{Various quantities of the orbit solution at different outburst epochs. The first column ($t_{\rm out}$) represents the starting time of outbursts in terms of the Julian year. The quantity $t_{\rm del}$ is the time delay between the impact and the outburst whereas $t_{\rm adv}$ stands for the time advance due to the bending of the accretion disk. 
The fifth column provides the dimensionless velocity of the secondary BH while $R_{\rm imp}$ denotes the distance from the center at which the impact occurs.
The next outburst is predicted to occur at end of July, 2019. \label{tab:orbit solution}}
\tablehead{
\colhead{$t_{\rm out}$(Julian year)} & \colhead{$t_{\rm del}(yr)$} & \colhead{$t_{\rm adv}(yr)$} & \colhead{$R_{\rm imp}(AU)$} & \colhead{$v_0/c$}}
\colnumbers
\startdata
1886.623 &   0.018  &    0.0    &   3837    & 0.251     \\
1896.668 &	 1.350	&	 0.176 	& 	15242 	& 0.088    	\\
1898.610 & 	 0.013	&	 0.0 	& 	3412	& 0.266    	\\
1906.196 & 	 2.882	&	 0.198 	& 	18384 	& 0.061    	\\
1910.592 &	 0.014	&	 0.0 	& 	3528	& 0.262 	\\
1912.978 &	 0.478	&	 0.104 	& 	11498 	& 0.121     \\
1922.529 &	 0.026	&	 0.0 	& 	4267 	& 0.238     \\
1923.725 &	 0.089	&	 0.052  & 	6589 	& 0.186 	\\
1934.335 &	 0.072	&	 0.0 	& 	6127 	& 0.194 	\\
1935.398 &	 0.028	&	 0.034 	&	4431 	& 0.233     \\
1945.818 &	 0.346	&	 0.0 	& 	10421 	& 0.131     \\
1947.282 &	 0.014	&	 0.027	& 	3540 	& 0.260     \\
1957.083 &	 2.254	&	 0.066	& 	17313 	& 0.067     \\
1959.212 &	 0.012	&	 0.026 	& 	3313 	& 0.267     \\
1964.231 &	 1.552	&	 0.060 	& 	15786 	& 0.079     \\
1971.126 &	 0.015	&	 0.028 	& 	3613 	& 0.255     \\
1972.928 &	 0.222	&	 0.0 	& 	8967 	& 0.146     \\
1982.964 &	 0.032	&	 0.037 	& 	4633 	& 0.224 	\\
1984.119 &	 0.049	&	 0.0 	& 	5387 	& 0.205     \\
1994.594 &	 0.110	&	 0.058	& 	7079 	& 0.173     \\
1995.841 &	 0.018	&	 0.0  	& 	3855 	& 0.245 	\\
2005.743 &	 0.631	&	 0.130 	& 	12427 	& 0.106 	\\
2007.693 &	 0.011	&	 0.0 	&	3259 	& 0.264     \\
2015.868 &	 2.392	&	 0.205 	& 	17566 	& 0.058     \\
2019.569 &	 0.011	&	 0.0	& 	3218 	& 0.265     \\
2022.548 &	 0.624	&	 0.131	& 	12386 	& 0.103     \\
2031.412 &	 0.016	&	 0.0 	& 	3708 	& 0.246     \\
2032.732 &	 0.103	&	 0.059	& 	6911 	& 0.170    	\\
2043.149 &	 0.041	&	 0.0	& 	5051 	& 0.207   	\\
2044.196 &   0.027  & 	 0.036 	& 	4409 	& 0.222    	\\
2054.591 &   0.170  & 	 0.0     & 	8197 	& 0.149     \\
2055.945 &	 0.012 	&	 0.028	& 	3352 	& 0.255     \\
\enddata
\end{deluxetable}

\subsection{Physical arguments for heuristically including the tail contributions to GW emission on our BBH dynamics} \label{subsec:tail_term_importance}

We are now in a position to explain why we were forced to introduce the $\gamma$ parameter and obtain an estimate for it from the impact flare timings. Recall that the prediction and analysis of the GR centenary flare observations were pursued using fully 3PN accurate orbital dynamics that incorporated the effects of dominant (Newtonian) order GW emission on BBH dynamics. Therefore, it is natural to extend the PN accuracy of our BBH dynamics by including the 3.5PN order contributions which are available in \cite{IW95}. However, it is not advisable to add only the 3.5PN order reactive contributions to BBH dynamics. This is because the fully 3.5PN order BBH equations of motion can provide extremely inaccurate secular orbital phase evolution during the inspiral regime of unequal mass compact binaries. The above statement is fully endorsed by the second column entries in Table~I of \cite{BDIWW}, which showed that the accumulated number of GW cycles (${\mathcal N}$) due to 1PN and 1.5PN tail contributions to GW emission tend to cancel each other for large mass-ratio binaries.
A close inspection of the above Table also reveals that the ${\mathcal N}$ estimate, based on fully 1PN-accurate orbital frequency evolution ($\dot \omega$), substantially increases the expected number of GW cycles for the orbital revolutions of unequal mass BH-NS binaries. 
Recall that 1PN accurate orbital frequency evolution corresponds to 3.5PN accurate orbital evolution in the PN description. These considerations are important for the BBH orbital evolution in OJ~287 as the fully 3.5PN accurate equations of motion can provide erroneous orbital phase evolution.
An erroneous orbital phase evolution ensures that the observed impact flares timings will not be consistent with the BBH central engine model.
Indeed, we have verified that the use of fully 3.5PN accurate equations of motion leads to a loss of acceptable solutions, and the situation is not improved by adding the 4.5PN order contributions. 
Therefore, it is crucial for us to incorporate the effect of hereditary contributions to GW emission on our BBH dynamics that appear at 4PN order in Equation~(\ref{eqn:eom}).
This is also influenced by the earlier mentioned fact that ${\mathcal N}$ estimates from 1PN contributions to $\dot \omega$ get essentially canceled by the 1.5PN `tail' contributions to $\dot \omega$ for unequal mass binaries.
For this reason, it is rather important for us to incorporate in our Equation~(\ref{eqn:eom}) terms that  can lead to 1.5PN accurate tail contributions in $\dot \omega$.

Therefore, we introduce a heuristic way of incorporating the effect of dominant-order tail contributions to GW emission into  BBH orbital dynamics.
This is essentially due to the nonavailability of 4PN(tail) contributions in the form of Equation~(\ref{eqn:rr_2p5-4p5}).
The plan, as noted earlier, is to introduce an ambiguity parameter $\gamma$ at the dominant-order radiation reaction terms such that the second line of Equation~(\ref{eqn:eom}) can be replaced using Equation~(\ref{eqn:gamma_tail}).
Note that we can now introduce an additional parameter while describing the dynamics of the BBH in our model as we have, at our disposal, the tenth fixed point from the timing of the November 2015  impact flare.
Indeed, the results we list in Tables~\ref{tab:parameters} and \ref{tab:orbit solution} are obtained by such a prescription for the BBH dynamics.

\begin{figure}
\plotone{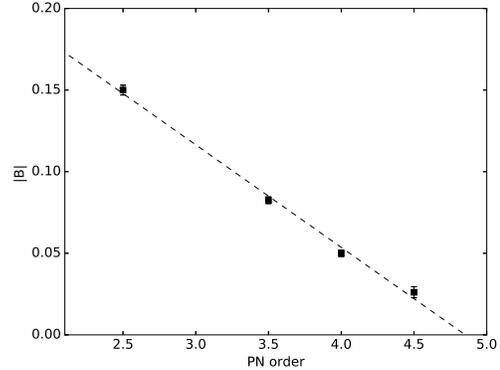}
\caption{A plot to demonstrate why the present order of PN approximation is sufficient to describe the secular evolution of BBH system in OJ~287. We plot appropriately scaled and dimensionless values of velocity component $B$ that appear at the reactive Newtonian and at 1PN, 1.5PN, and 2PN orders. The $B$ coefficients are chosen as they drive the secular evolution of BBH binary in OJ~287. The inferred linear regression suggests that further higher order PN contributions should  not be relevant in our model.
\label{fig:radreaction}}
\end{figure}

 Clearly, our procedure for evolving the BBH orbit  requires further scrutiny.
It is natural to ask if additional higher order radiation reaction contributions to $\ddot{\vek x}$ are required while evolving the BBH orbit in OJ~287.
We gather from various numerical experiments, associated with the above detailed orbit-fitting procedure, that the velocity-dependent terms in  Equation~(\ref{eqn:eom}) are crucial for incorporating the effects of GW emission on the dynamics of BBH system in OJ~287. 
A plot of appropriately scaled and dimensionless $B$ coefficients that appear at 
2.5PN, 3.5PN, 4PN, and 4.5PN orders in Equation~(\ref{eqn:rr_2p5-4p5}) is displayed in  Figure~\ref{fig:radreaction}.
The visible linear regression suggests that the further higher-order contributions to GW emission would not substantially influence the orbital evolution of the BBH in OJ~287.
Note that the contribution appearing at 4PN order in Figure~\ref{fig:radreaction} is obtained by multiplying the scaled dimensionless 2.5PN order B coefficient by $0.304$,which arises by subtracting unity from the our $ \gamma_{obs} = 1.304$ estimate.

 In the next section, we show that the extracted $\gamma_{obs}$ value is consistent with the general relativistic orbital phase evolution of the BBH in OJ~287 that  explicitly incorporates the effects of dominant-order tail contributions to GW emission.

\section{Estimating $\gamma$ from General Relativistic Considerations} \label{sec:radfac}

To obtain a GR-based estimate for $\gamma$ (we call it $\gamma_{GR}$), we adapted the approach that provided the accurate temporal orbital phase evolution for compact binaries inspiraling along PN-accurate eccentric orbits \citep{DGI,KG06}.
This GW phasing approach is required to construct both time and frequency domain templates to model inspiral GWs from compact binaries in eccentric orbits in GR \citep{THG}.
The approach ensures accurate BBH orbital phase evolution, as the success of {\it matched filtering}, employed to extract weak GW signals from noisy interferometric data, demands GW templates with accurate phase evolution \citep{SS09}.
This feature is crucial for the present problem, too, as accurate predictions of impact flare timings demand accurate knowledge about the orbital phase evolution of the BBH in OJ~287.
In other words, the OJ~287 impact flares occur when the secondary BH crosses the accretion-disk plane of the primary BH at constant phase angles of the orbit like $0, \pi, 2\pi, 3\pi, \dots$, and therefore the accurate determination of the BBH orbital phase evolution should lead to precise predictions for the impact flare timings.
This is why we are adapting the GW phasing approach to determine a GR-based estimate for $\gamma$ ($\gamma_{GR}$).

\subsection{GW phasing formalism}

In GW phasing approach, we are interested in accurate evolution of the BBH orbital phase which takes into account the effects of the conservative and reactive terms present in first and second line of Equation~(\ref{eqn:eom}) respectively. This approach accurately incorporate the effects of first two lines of Equation~(\ref{eqn:eom}) without explicitly invoking them. In our implementation, we employ a 3PN-accurate Keplerian-type parametric solution to model the conservative parts of the orbital dynamics, i.e., to model the first line of Equation~(\ref{eqn:eom}) \citep{MGS}. This allows us to express the temporally evolving BBH orbital phase $\phi$ as
\begin{equation}
\phi - \phi_0 = ( 1 + k ) \, l \,,
\label{eqn:phase_evolution}
\end{equation}
where $l$ is the mean anomaly defined to be $ l = 2\, \pi ( t -t_0)/P_b$ \citep{MGS,KG06}.
%{\new (we will use the mean anomaly $l$ in place of time $t$ to track the evolution of orbital phase)}.
%http://adsabs.harvard.edu/abs/2004PhRvD..70j4011M
Initial values of the orbital phase and its associated coordinate time are denoted by $\phi_0$ and $t_0$, while $P_b$ stands for the radial orbital period of a PN-accurate eccentric orbit.
Furthermore, the dimensionless fractional periastron advance per orbit $k$ ensures that we are dealing with a precessing eccentric orbit.
A close comparison with \cite{KG06} reveals that we have neglected 3PN-accurate periodic contributions to the orbital phase in the above equation. This is justified as we are focusing our attention on the secular evolution of the BBH orbital phase. 
The fractional rate of periastron advance $k$ is an explicit function of $n$ and $e_t$ where $n =2\, \pi/P_b$ is the mean motion and $e_t$ provides the eccentricity parameter that enters the PN-accurate Kepler equation of \cite{MGS}. The 3PN-accurate expression for $k$ is given in the Appendix~\ref{app:kdndldetdl}. 
Note that it is by using the 3PN-accurate expression for $k$ that we are incorporating the BBH orbital phase evolution at the 3PN-accurate conservative level into $\ddot{\vek x}$.

 The next step requires us to model the influences of the reactive terms (second line of Equation~(\ref{eqn:eom})) on the conservative 3PN-accurate orbital phase evolution.
This is done by providing differential equations that describe temporal evolutions for $n$ and $e_t$ due to emission of GWs, consistent with the ``reactive" PN orders of Equation~(\ref{eqn:eom}); details are provided by \cite{KG06}.
Following \cite{BDI}, it is convenient to split the fully 2PN-accurate differential equations for $n$ and $e_t$ into two parts, given by
\begin{subequations}
\begin{eqnarray}
\left( \frac{ d n }{ dl } \right)^{\rm {2PN}} &=& \left( \frac{ d n }{ dl } \right)^{\rm Inst} + \left( \frac{ d n }{ dl } \right)^{\rm Tail}\,,\\
\left( \frac{ d e_t }{ dl } \right)^{\rm 2PN} &=& \left( \frac{ d e_t }{ dl } \right)^{\rm Inst} + \left( \frac{ d e_t }{ dl } \right)^{\rm Tail}\,,
\end{eqnarray}
\end{subequations}
where we used $ n\, dt = dl $ to obtain $(dn/dl,de_t/dl)$ expressions from their $(dn/dt,de_t/dt)$ counterparts of \cite{KG06}.
We term those contributions that depend only on the state of the binary at the usual retarded instant $T_{\text r}$ as the ``instantaneous" contributions and refer those terms that are {\it a priori} sensitive to the BBH dynamics at all previous instants to $T_{\text r}$ as the ``tail" (or hereditary) terms.
Moreover, these instantaneous contributions appear at the ``absolute" 2.5PN, 3.5PN, and 4.5PN orders like the reactive contributions in Equation~(\ref{eqn:eom}) for $\ddot{\bf x}$, while the ``tail" contribution enters at the ``absolute" 4PN order.
Therefore, the  differential equations for evolution of $n$ and $e_t$ may also be symbolically  written as
\begin{subequations}
\begin{eqnarray}
\left( \frac{ d n }{ dl } \right)^{\rm eaxct}_{\rm  2PN} &=& 
\left( \frac{ d n }{ dl } \right)_{\rm 2.5PN} 
+ \left( \frac{ d n }{ dl } \right)_{\rm 3.5PN} \nonumber \\ &&
+ \left( \frac{ d n }{ dl } \right)_{\rm 4.5PN} 
+\left( \frac{ d n }{ dl } \right)_{\rm 4PN}\,, \label{eqn:dndl_ext}
\\
\left( \frac{ d e_t }{ dl } \right)^{\rm exact}_{\rm 2PN} &=& 
\left( \frac{ d e_t }{ dl } \right)_{\rm 2.5PN} 
+ \left( \frac{ d e_t }{ dl } \right)_{\rm 3.5PN} \nonumber \\ &&
+  \left( \frac{ d e_t }{ dl } \right)_{\rm 4.5PN}\,
+ \left( \frac{ d e_t }{ dl } \right)_{\rm 4PN}\,.\label{eqn:detdl_ext}
\end{eqnarray}
\end{subequations}
The explicit expressions for these 2PN-accurate contributions are listed in 
Appendix~\ref{app:kdndldetdl}.

A few comments are in order.
Note that these orbital-averaged equations are derived by computing time derivatives of 2PN-accurate $n$ and $e_t$ expressions in the {\it modified harmonic gauge}, available in \cite{MGS}.
We apply the heuristic arguments that the  time derivatives of the binary orbital energy and angular momentum should be balanced by their far-zone GW energy and angular momentum fluxes.
Close inspection of \cite{KG06} and the above equations reveals that there exists no strict one-to-one correspondence between different PN orders present in the second line of Equation~(\ref{eqn:eom}) and the above equations for $n$ and $e_t$.
In other words, 3.5PN-order contributions to the differential equations arise from both 2.5PN and 3.5PN terms in Equation~(\ref{eqn:eom}), and this will be relevant for us while estimating the value $\gamma_{GR}$.
Furthermore, it is the use of certain rational functions of $e_t$ that allowed us to write closed-form expressions for the tail contributions to $dn/dl$ and $d e_t/dl$, as explained by \cite{THG}.

\subsection{Estimation of $\gamma_{GR}$ from GW phasing}

We can now obtain secular orbital phase evolution of the BBH in OJ~287 due to the action of fully 2PN-accurate reactive dynamics on the fully 3PN-accurate conservative dynamics.
It should be noted that this is what the first and second-line contributions in Equation~(\ref{eqn:eom}) aim to provide while implementing our BBH central engine model, provided we ignore periodic contributions to its solution.
We obtain the desired orbital phase evolution by numerically imposing on $ \phi - \phi_0 = ( 1+ k)\,l$, given by Equation~(\ref{eqn:phase_evolution}), the secular evolutions in $n(l)$ and $e_t(l)$ due to Equations~(\ref{eqn:dndl_ext}) and (\ref{eqn:detdl_ext}). The resulting phase evolution is treated as $\phi_{\rm exact}$ (as we use exact 2PN accurate equations for $dn/dl$ and $de_t/dl$) in our effort to obtain $\gamma_{GR}$ associated with the BBH in OJ~287. 
Let us emphasize that the use of fully 2PN-accurate differential equations for $n(l)$ and $e_t(l)$ ensures that the orbital phase evolution does incorporate the effects of the dominant-order hereditary contributions to the GW emission.

To obtain $\gamma_{GR}$, we repeat the above procedure while employing slightly different differential equations for $n$ and $e_t$ that do not contain the tail contributions. 
This is expected as we are trying to model the BBH orbital phase evolution defined by the first line of Equation~(\ref{eqn:eom}), while the second-line contributions are replaced by Equation~(\ref{eqn:gamma_tail}) to model the reactive contributions to $\ddot {\bf x}$.
In other words, the instantaneous contributions to $dn/dl$ and $de_t/dl$ are modified to include the effect of the $\gamma$ factor, present in  Equation~(\ref{eqn:gamma_tail}).
The resulting Newtonian (quadrupolar {or 2.5PN}) contributions to $dn/dl$ and $de_t/dl$ are multiplied by the $\gamma_{GR}$ factor. 
However, the 3.5PN-order Instantaneous contributions now consist of two parts. The first part contains $\gamma_{GR}$ as a common factor and arises from the 2.5PN terms in $\ddot{\mathbf{x}}$. The second part is independent of $\gamma_{GR}$. 
The resulting differential equations for $n$ and $e_t$ can be written as 

\begin{subequations}
\centering
\label{eqn:dnet_approx}
\begin{eqnarray}
%\label{gamma_inst_2}
\left( \frac{ d n }{ dl } \right)^{\rm approx}_{\rm  2PN} &=& 
\gamma_{GR}\,\left( \frac{ d n }{ dl } \right)_{\rm 2.5PN} + 
\gamma_{GR} \left( \frac{ d n }{ dl } \right)_{\rm 3.5PN}^{(\rm 2.5PN)} \nonumber \\ &&
+ \left( \frac{ d n }{ dl } \right)_{\rm 3.5PN}
+ \left( \frac{ d n }{ dl } \right)_{\rm 4.5PN}\, , \label{eqn:dndl_approx}
\\
\left( \frac{ d e_t }{ dl } \right)^{\rm approx}_{\rm 2PN} &=& 
\gamma_{GR} \, \left( \frac{ d e_t }{ dl } \right)_{\rm 2.5PN} 
+ \gamma_{GR} \, \left( \frac{ d e_t }{ dl } \right)_{\rm 3.5PN}^{\rm (2.5PN)} \nonumber \\ &&
+ \left( \frac{ d e_t }{ dl } \right)_{\rm 3.5PN}
+  \left( \frac{ d e_t }{ dl } \right)_{\rm 4.5PN}\, ,\label{eqn:detdl_approx}
%+ \left( \frac{ d e_t }{ dl } \right)^{\rm Tail}\,,
\end{eqnarray}
\end{subequations}
where the 2.5PN, 3.5PN, and 4.5PN terms stand for the relative Newtonian, 1PN, and 2PN contributions due to GW emission, and the influence of the tail contributions needs to be captured by certain values of $\gamma$ for the BBH in OJ~287.
The appearance of $\gamma_{GR}$ at 2.5PN and 3.5PN orders in Equations~(\ref{eqn:dnet_approx}) is due to the introduction of  $\gamma_{GR}$ at the 2.5PN order in Equation~(\ref{eqn:eom}). This is because the time derivatives of PN-accurate orbital energy and angular momentum using Equation~(\ref{eqn:gamma_tail}) are required to obtain Equations~(\ref{eqn:dnet_approx}) as detailed in \cite{KG06}.
We do not list explicitly these contributions, and as expected these contributions add to the instantaneous contributions in $ dn/dl$ and $de_t/dl$, given by Equations~(\ref{eqn:dndl_ext}) and (\ref{eqn:detdl_ext}), when we let $\gamma=1$.
However, it is straightforward to modify Equations~(28)--(33) of \cite{KG06} to obtain $\gamma$ versions of Equations~(32) of  \cite{KG06},
and they provide the 2.5PN and 3.5PN contributions to our Equations~(\ref{eqn:dnet_approx}).
As to the  4.5PN contributions to the above-listed approximate  equations for
$ dn/dt$ and $de_t/dt$, we employed 2PN contributions in Equations~(B2) and (B3) of \cite{KG06}.

We are now in a position to repeat the numerical procedure that gave us $\phi_{\rm exact}$ while employing Equations~(\ref{eqn:dndl_approx}) and (\ref{eqn:detdl_approx}) for $ dn/dl$ and $de_t/dl$. The resulting accumulated orbital phase evolution is termed as $\phi_{\rm approx}$ (as we are using approximate formula for $dn/dl$ and $de_t/dl$ which involves $\gamma_{GR}$). 
To obtain $\gamma_{GR}$, we demand that $\Delta \phi = \phi_{\rm exact} - \phi_{\rm approx}$ should be smaller than $(\Delta \phi)_{\rm tol}$: an observationally extracted tolerable value for the accumulated orbital phase for the BBH in OJ~287.
This is justified as uncertainties in the observed outburst timings constrain the accuracy with which we can track the orbital phase evolution of the BBH in OJ~287.
An estimate for $(\Delta \phi)_{\rm tol}$ is obtained by noting that at best the uncertainty in the observed outburst timing is roughly 0.0005 yr.
In the BBH model, the associated  minimum $\phi$ uncertainty  occurs at the apocenter as the secondary moves slowly there.
Invoking  a Keplerian orbit and the expression for $d \phi/dt$ at the apocenter, we write 
\begin{eqnarray*}
\frac{d \phi}{dt} &=& \frac{2\pi}{P_{\rm orb}} \frac{\sqrt{1-e^2}}{(1+e)^2}\\
&=&\frac{2\pi \times 0.263 }{P_{\rm orb}}\,.
\end{eqnarray*}
This leads to an observationally relevant $(\Delta \phi)_{\rm tol}$ estimate as 
\begin{equation}
(\Delta \phi)_{\rm tolerance} = \frac{0.0005 \times 2 \pi \times 0.263}{9.385} = 0.000088.
\end{equation}
We have checked that the inclusion of the PN corrections to $d \phi/dt$ does not substantially vary the above estimate for $(\Delta \phi)_{\rm tol}$.

To estimate $\gamma_{GR}$, we equate the earlier estimated difference in the accumulated BBH orbital phase $\Delta \phi$ to $(\Delta \phi)_{\rm tol}$ while considering roughly 12 orbital cycles of the BBH in OJ~287. 
This number of orbital cycles roughly corresponds to the span of observational data on OJ~287 ( $\sim 130$ yr) that we used to determine $\gamma_{obs}$ from impact flare timings.
In practice, we let $(\Delta \phi)_{\rm tol }$ vary between $-10^{-4}$ and $+\,10^{-4}$ while varying $\gamma_{GR}$ during  $(\Delta \phi)$ estimations.
We infer that $(\Delta \phi)_{\rm tol }$ forces the $\gamma_{GR}$ estimates to be $1.2917 \pm 0.0045$.
This is a very encouraging result as our GR-based estimate for $\gamma$ ($\gamma_{GR}$) is fairly close to our earlier estimate $\gamma_{obs}$, listed in Table~\ref{tab:parameters}, that was purely based on the {\it observed} impact flare timings while employing Equation~(\ref{eqn:gamma_tail}) to model the reactive contributions to $\ddot {\bf x}$.

\begin{figure}
\plotone{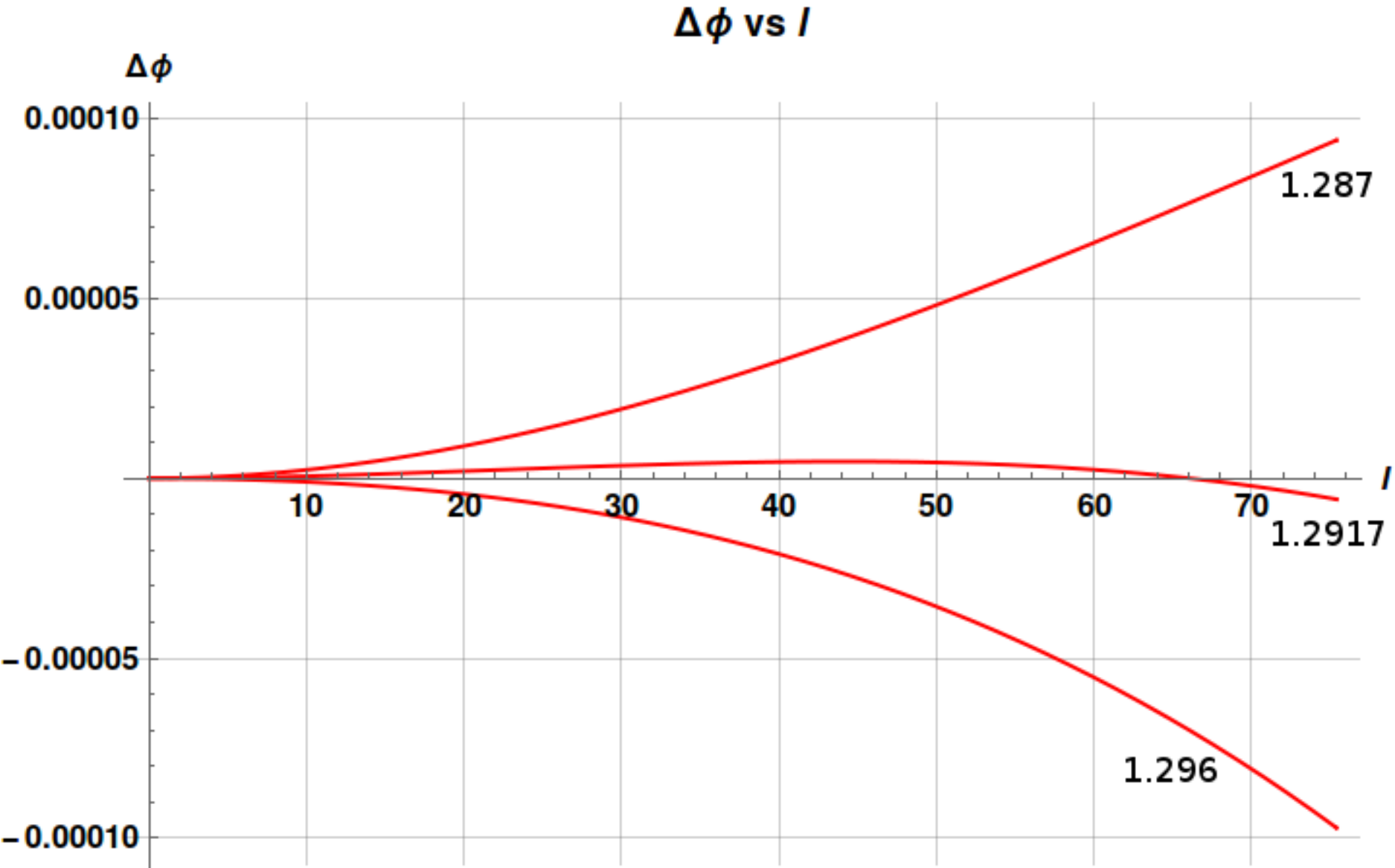}
\caption{ Plots of $\Delta\phi$ in radians as a function of mean anomaly $l$ for different $\gamma_{GR}$ values, for an $l$ interval corresponding to 12 orbital cycles of the OJ~287 BBH system. This $l$ value roughly corresponds to the time interval for which we have optical data on OJ~287.
We observe that for $\gamma_{GR}$ values between $1.287$ and $1.296$, the $\Delta \phi$ estimates are smaller than the present estimate for $\Delta\phi_{\rm tol}$.} 
\label{fig:radfac}
\end{figure}

We have verified that the inclusion of the spin-orbit contributions to the PN-accurate orbital phase evolution does not affect the above estimate.
In Figure~\ref{fig:radfac} we plot $\Delta\phi (l) $ as a function of $l$ for different $\gamma_{GR}$ values to show its secular variations.
These plots show that the differences between our two (exact and approximate) estimates for the accumulated BBH orbital phase remain within the $(\Delta \phi)_{\rm tol }$ values while varying $\gamma_{GR}$ between $1.287$ and $1.296$.
Note that the quantity `$\gamma_{GR}-1$' provides the present comparative strength of the dominant order tail contributions to BBH dynamics in comparison with the quadrupolar order gravitational radiation reaction terms that appear at 2.5PN order.
The expected GW emission induced hardening of BBH orbit ensures that the effects of higher order contributions like the tail terms can be more prominent during later epochs, which implies that the value of $\gamma$ will be epoch dependent.
The present detailed analysis opens up the possibility of evolving the BBH in OJ~287 with the help of Equations~(\ref{eqn:eom}) and (\ref{eqn:gamma_tail}) while using the  GR-based $\gamma$ estimate obtained above. 
In the next section, we explore the observational benefits of such an approach after taking a careful look at the light curves of OJ~287 in the vicinity of impact flare epochs.

\section{Predicting the Optical Light Curve of OJ~287 Near the Next Impact Flare Epoch} \label{sec:lc_comparison}

The present section explores our ability to model the expected optical light curve of OJ~287 during the next impact flare epoch.
This is attempted by comparing historical observational datasets with what we expect from our model while heavily depending on data from the 2015--2017 observational campaign on OJ~287. 
We also explain the astrophysical reward of predicting the impact flare light curve around the next impact outburst and its actual observations.

\subsection{ Optical Light Curves of OJ~287: Observations and Model Comparisons}

%\iffalse

\begin{figure*}
\gridline{\fig{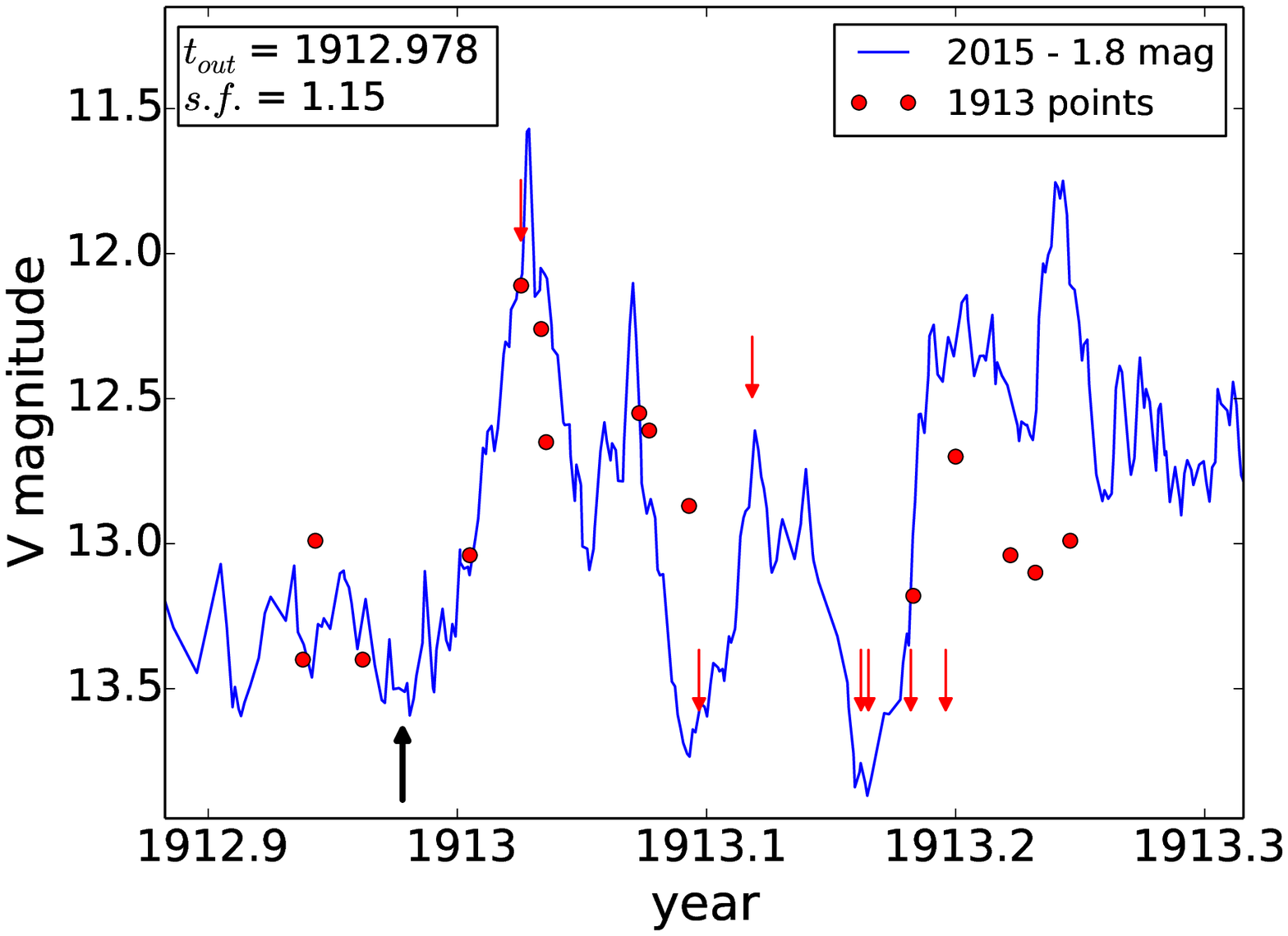}{0.33\textwidth}{(a) correlation = 0.82}
          \fig{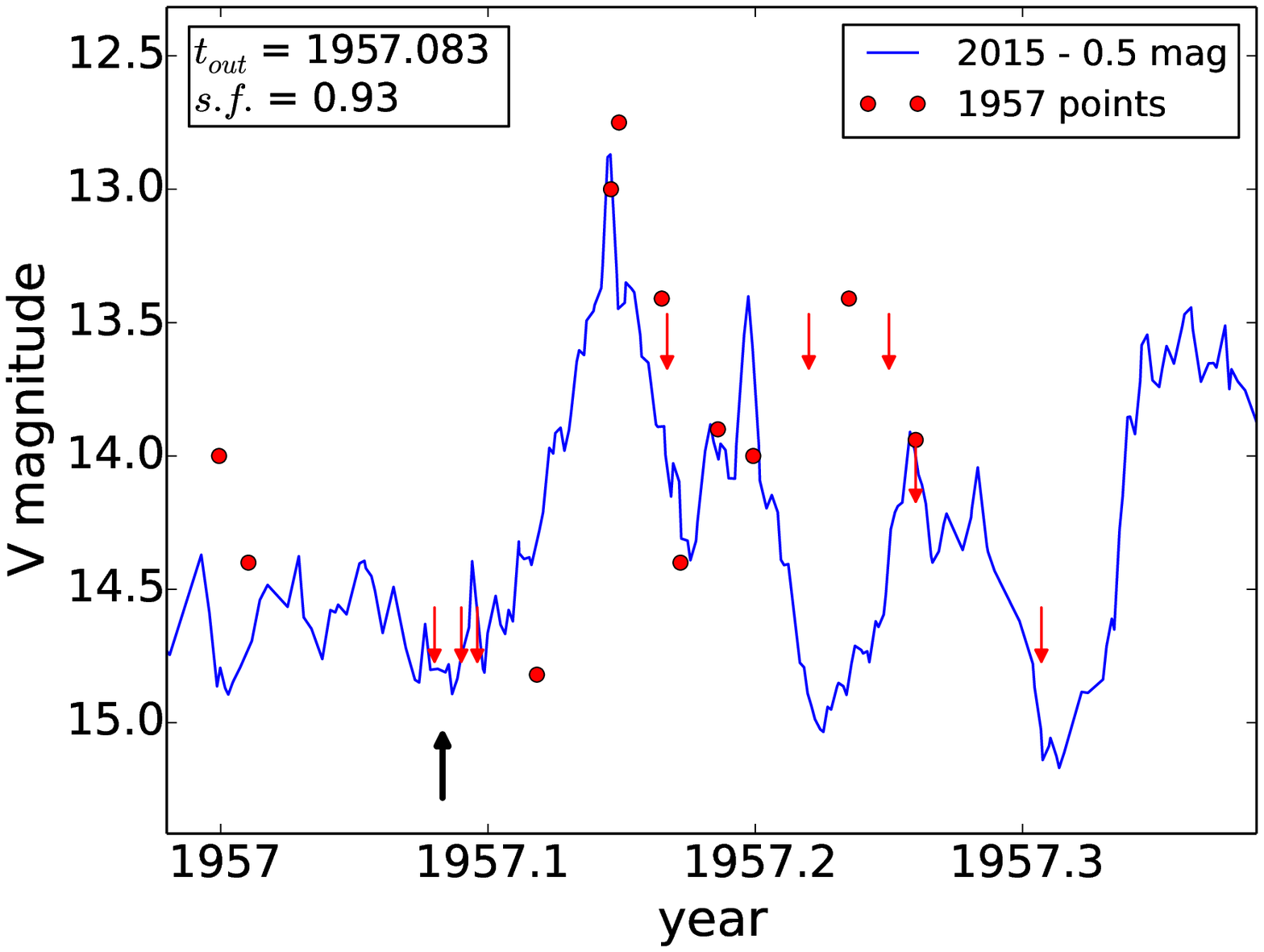}{0.33\textwidth}{(b) correlation = 0.89}
          }
\gridline{\fig{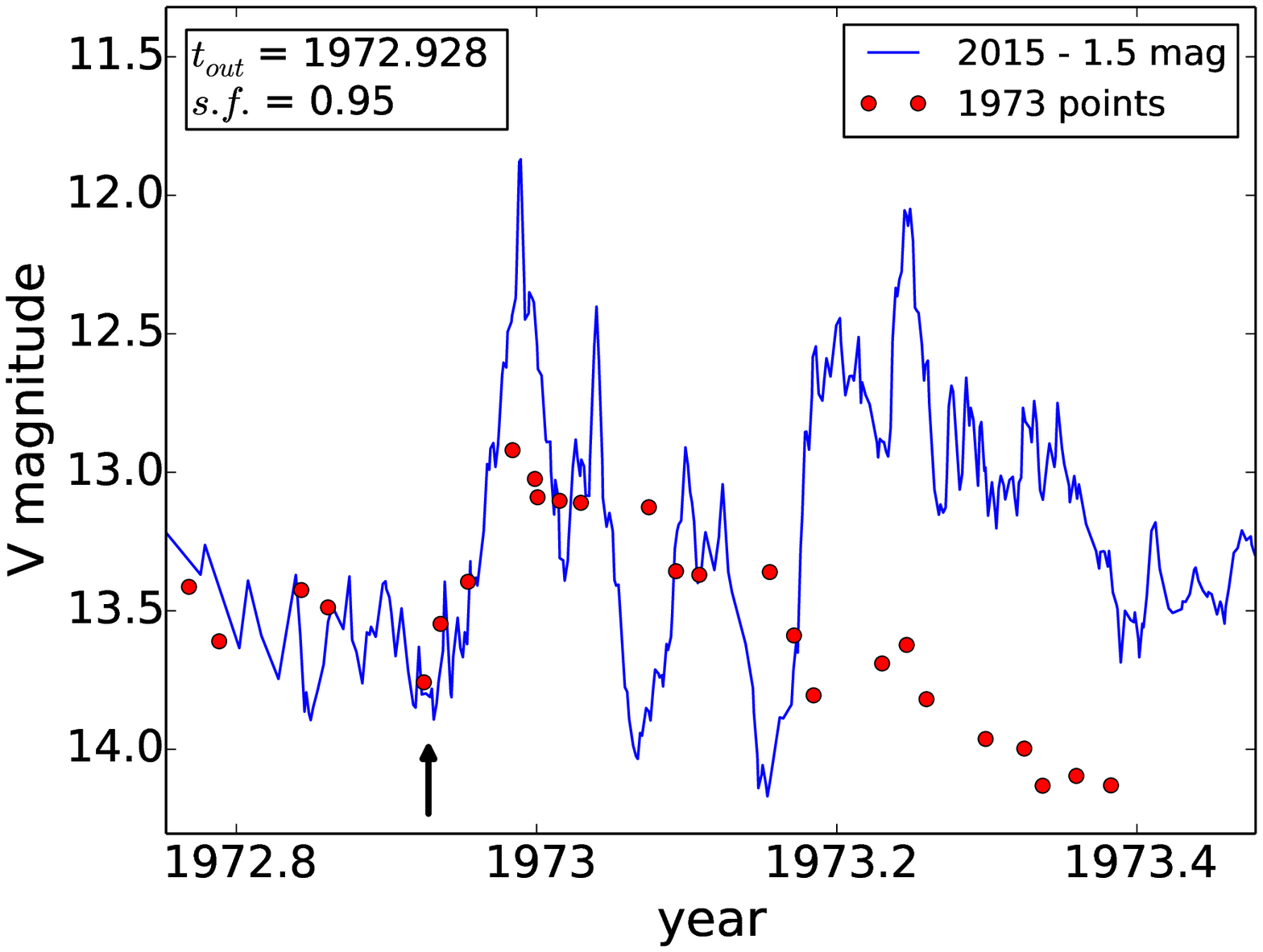}{0.33\textwidth}{(c) correlation = 0.87}
		  \fig{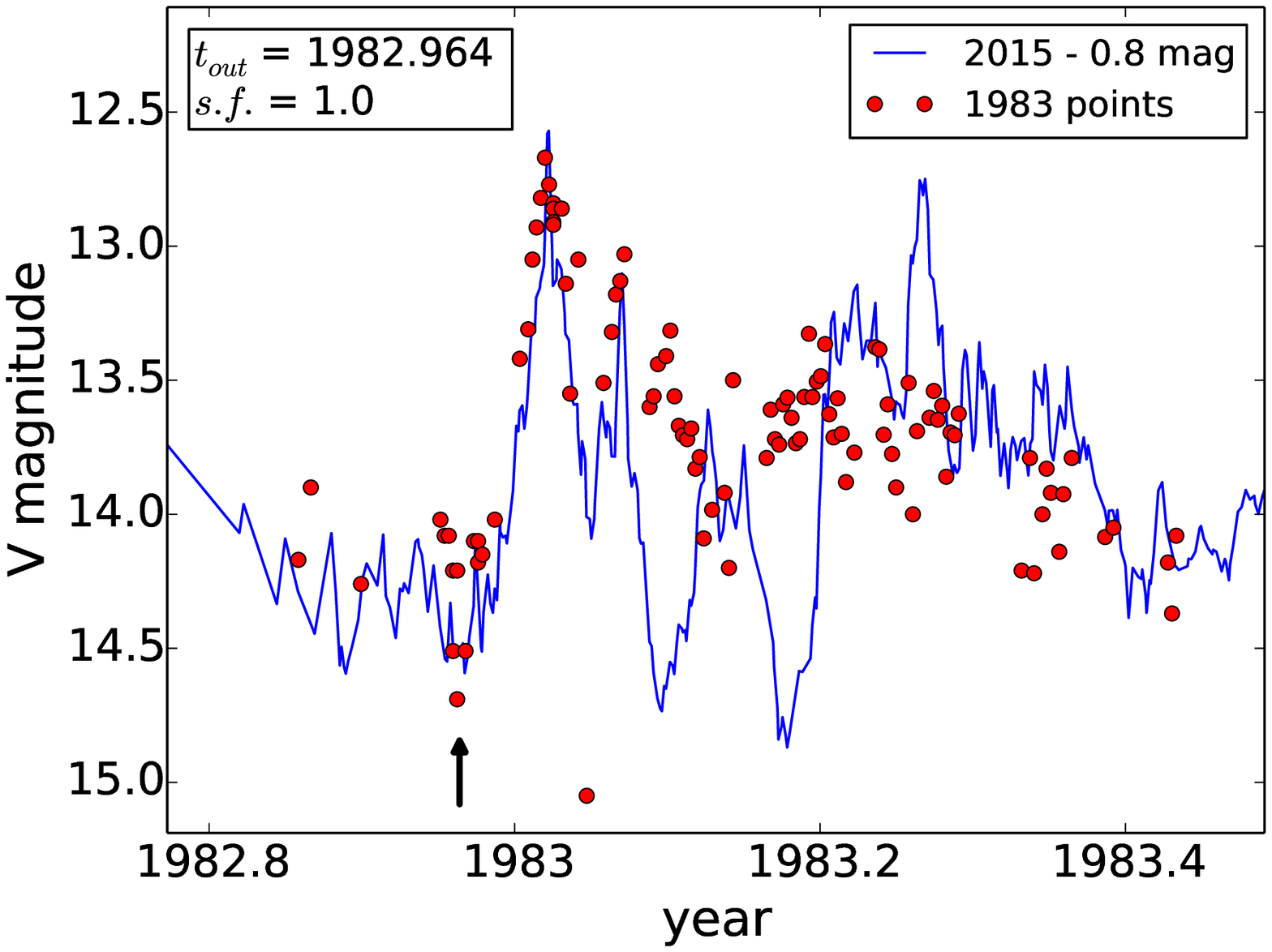}{0.33\textwidth}{(d) correlation = 0.88}
          }
\gridline{\fig{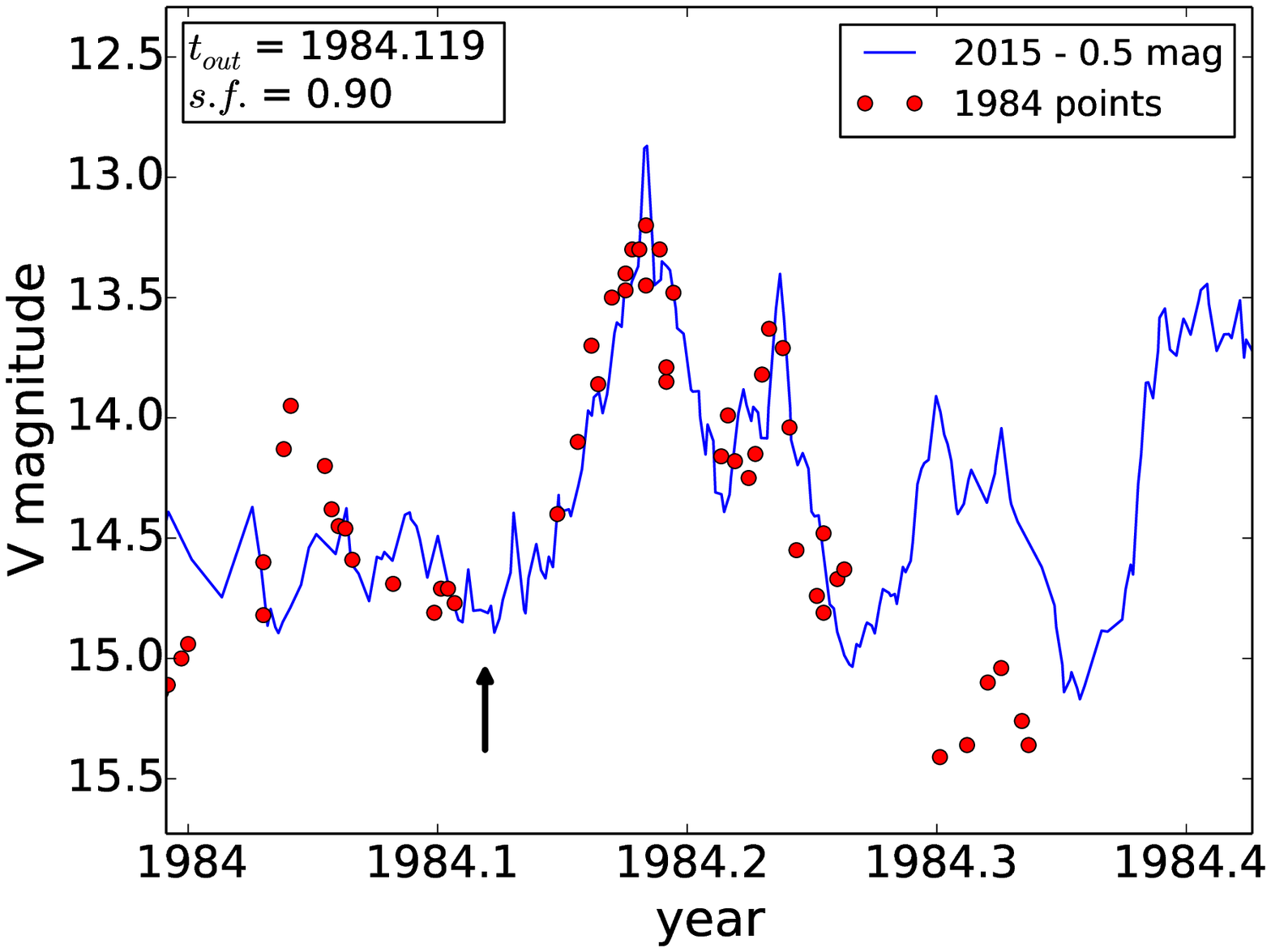}{0.33\textwidth}{(e) correlation = 0.91}
		  \fig{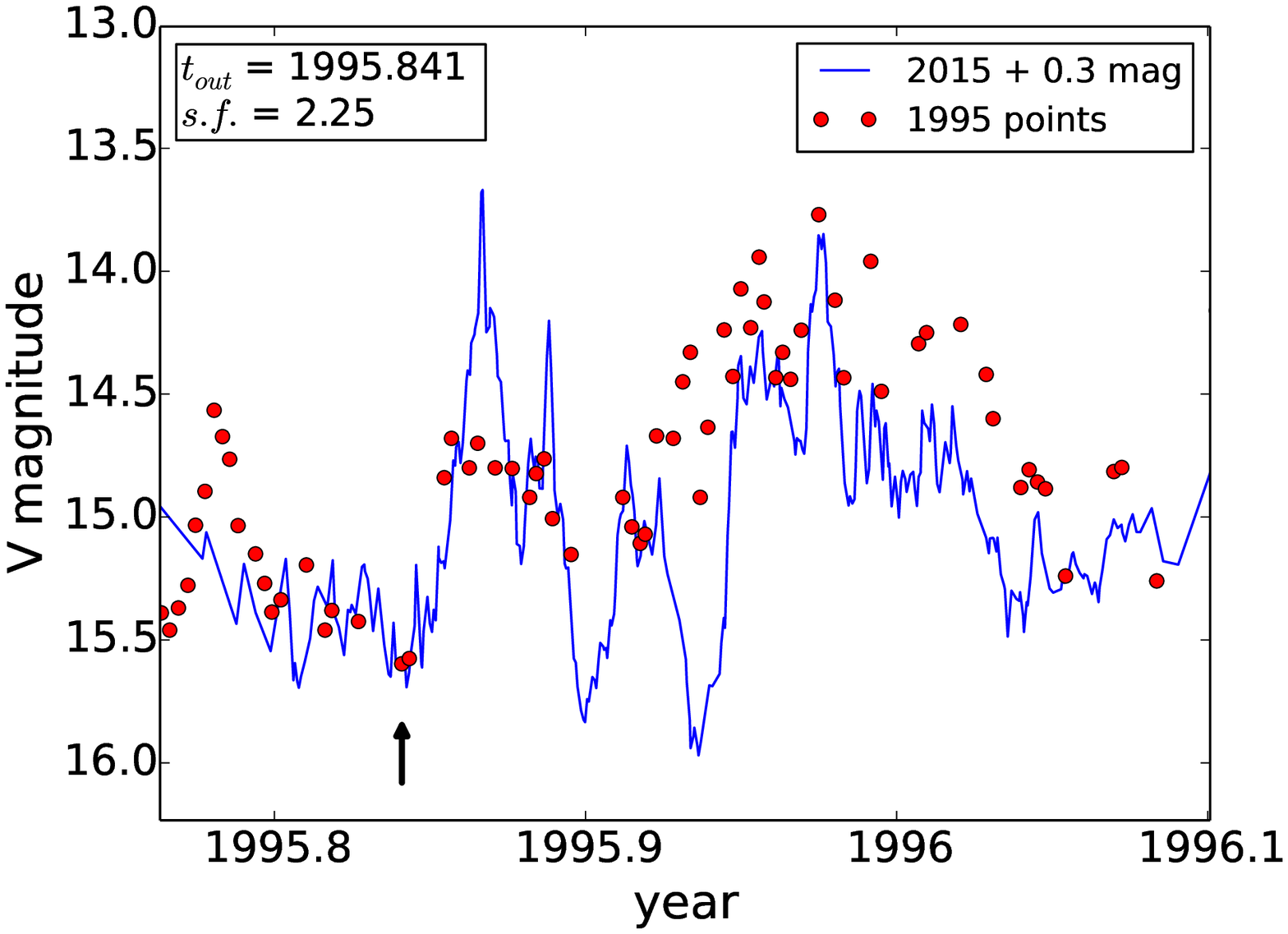}{0.33\textwidth}{(f) correlation = 0.82}
          }
\caption{Light-curve comparisons of well-observed outbursts. Tips of the red thin arrows (pointing downward) indicate upper limits, whereas the black thick arrows (pointing upward) indicate the starting time of outbursts, represented by $t_{\rm out}$. The 2015 template light curve is either stretched (if $s.f. < 1.0$) or squeezed (if $s.f. > 1.0$) depending on the speed factor ($s.f.$). The correlations of outburst light curves with the templates are given below the figures. The correlations have been calculated for a time interval of 2 months (in the 2015 timescale) around the outburst (using Mathematica). The high correlations indicate that the shapes of the outburst light curves (if the $s.f.$ is taken into account) are similar. \label{fig:correlation}}
\end{figure*}

\begin{figure*}
\gridline{\fig{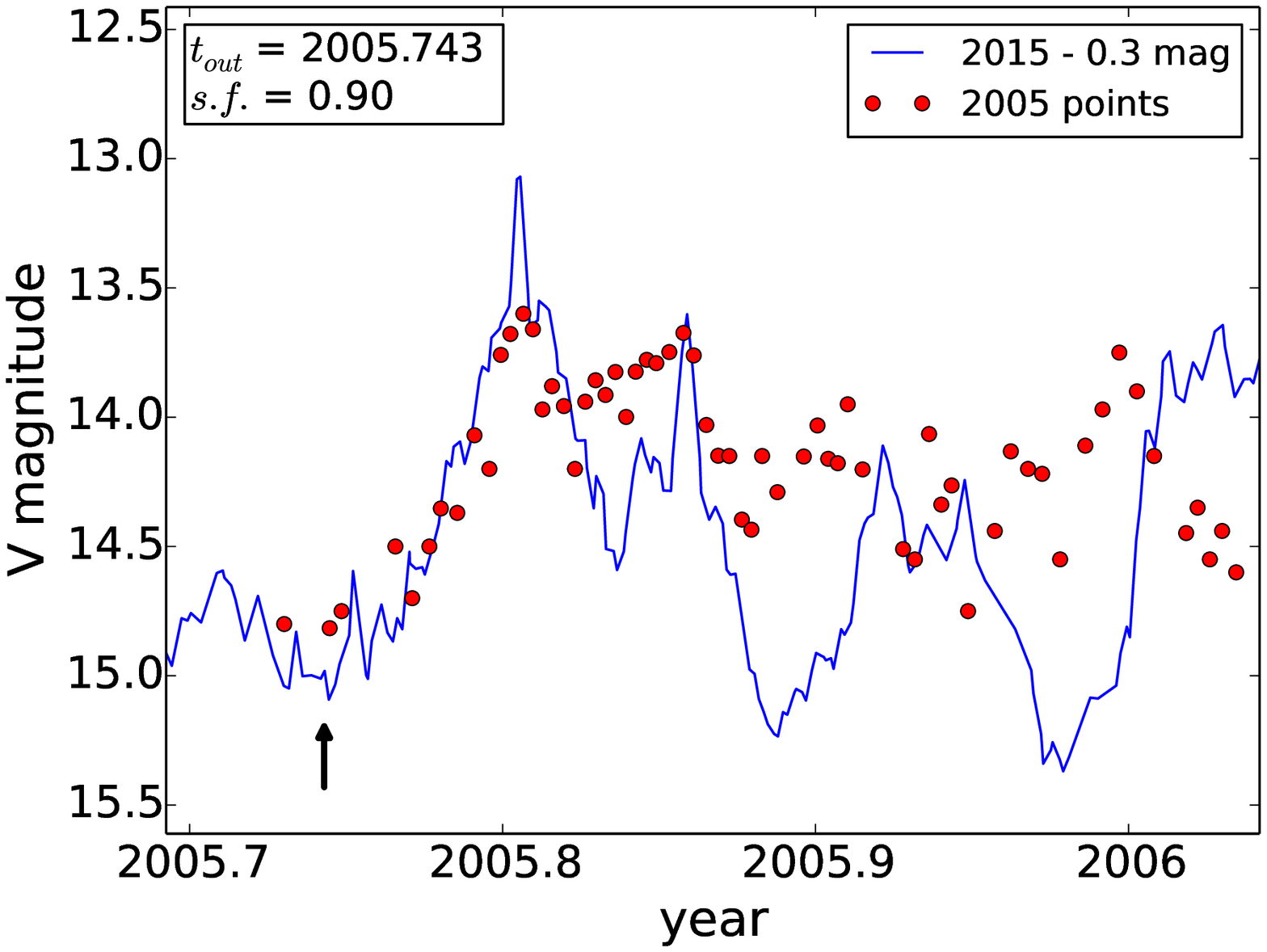}{0.33\textwidth}{(a) correlation = 0.84}
		  \fig{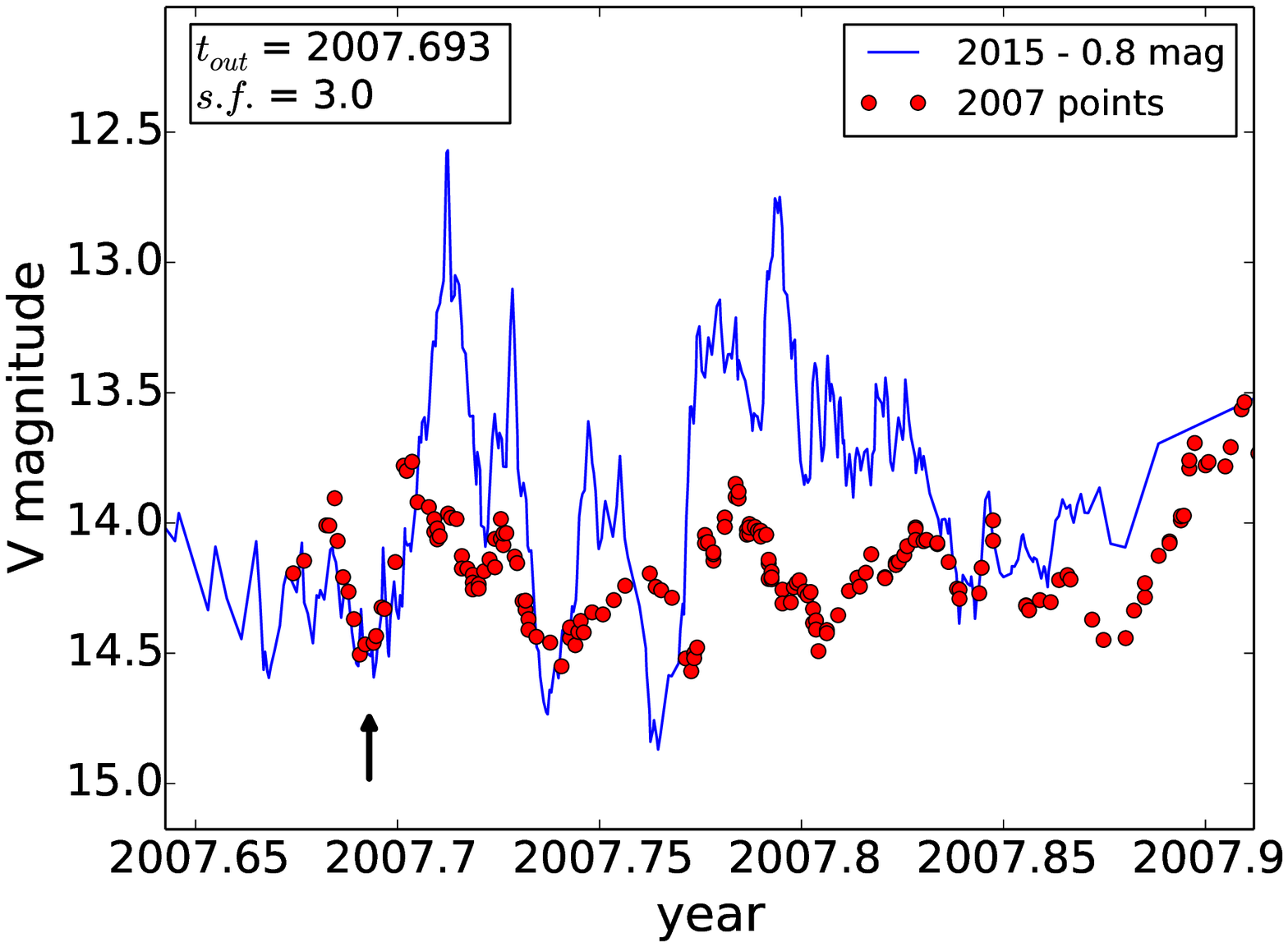}{0.33\textwidth}{(b) correlation = 0.65}    
          }
\gridline{\fig{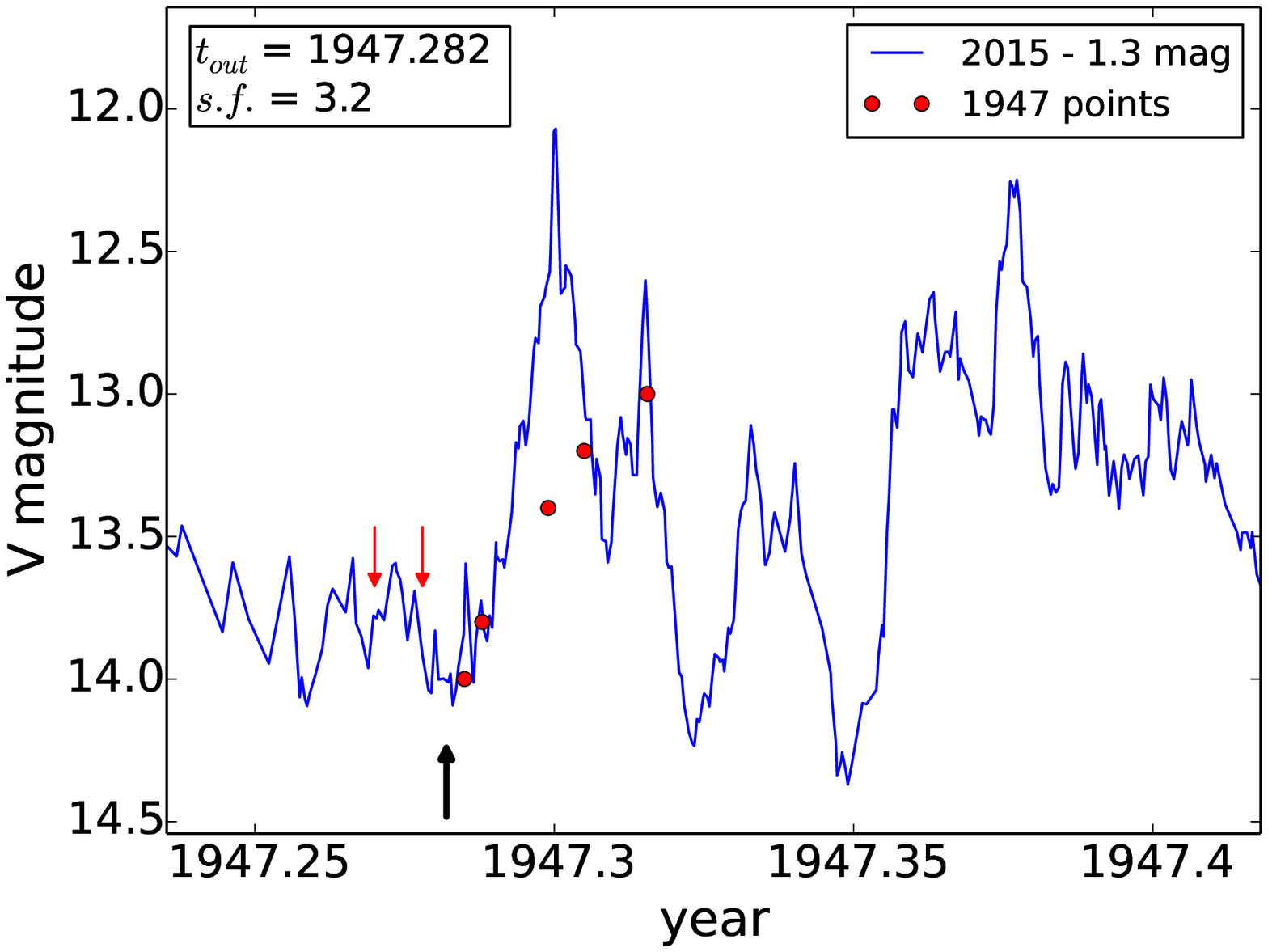}{0.33\textwidth}{(c)}
		  \fig{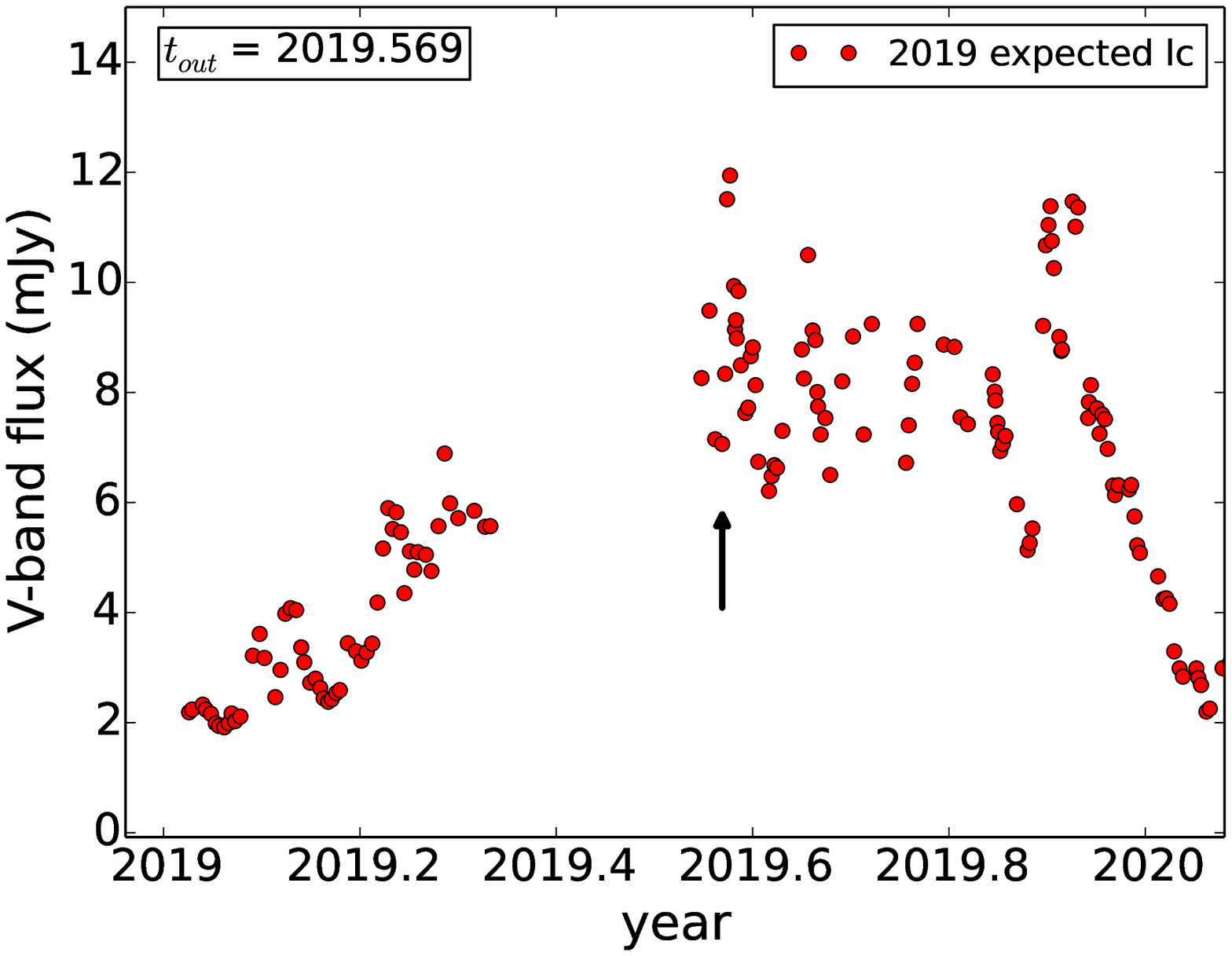}{0.33\textwidth}{(d)}
          }
\caption{Panels (a)--(c): Respectively, light-curve comparisons of the 2005, 2007, and 1947 outbursts with the 2015 template. The correlation for the 2007 outburst (panel (b)) is low because the rise in magnitude for the 2007 outburst peak is low. For the 1947 outburst (panel (c)), we do not have enough data points to calculate the correlation. In panel (d) we show the predicted light curve for the 2019 July outburst. \label{fig:correlation1}}
\end{figure*}

\begin{figure*}
\gridline{\fig{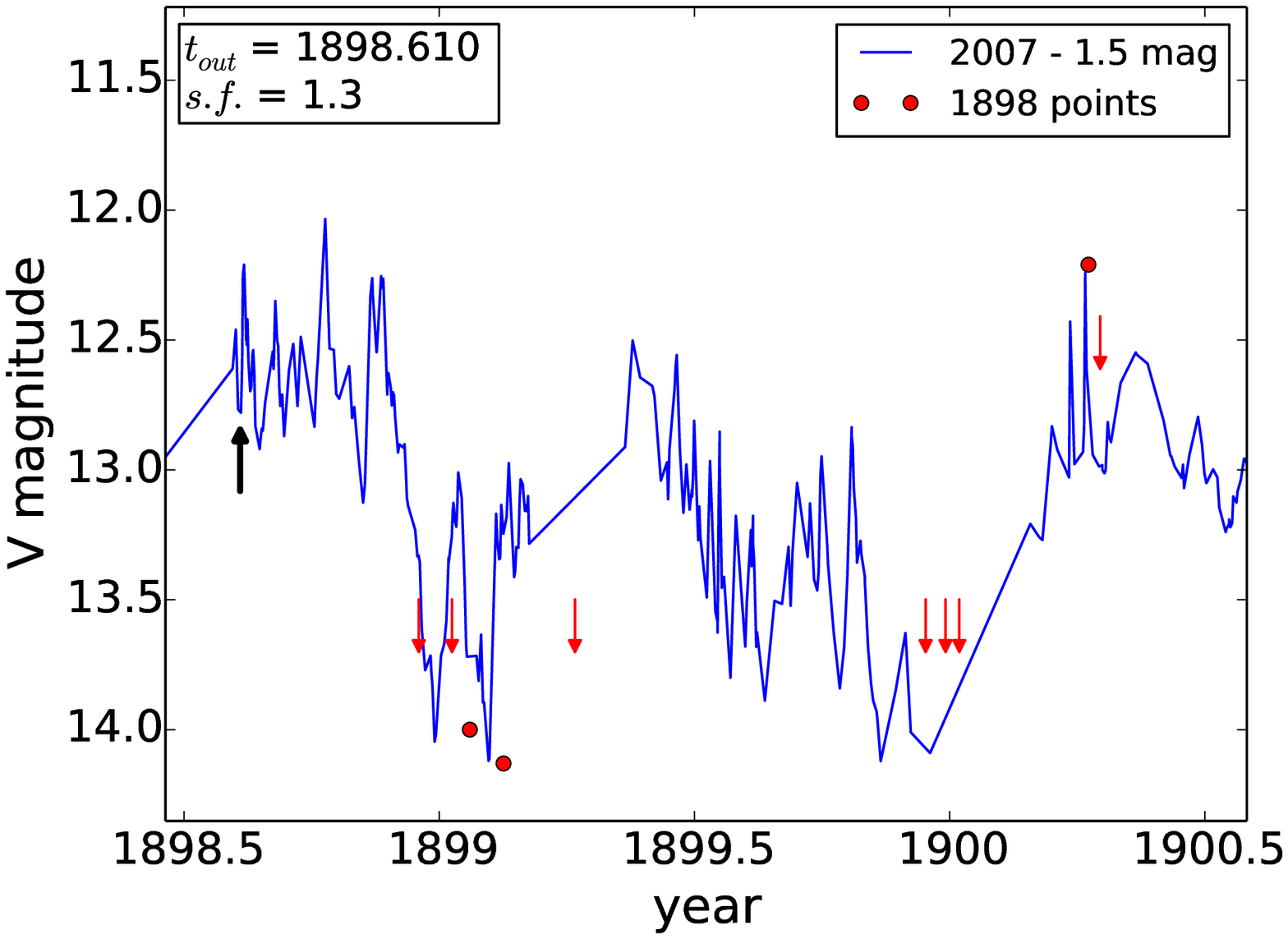}{0.33\textwidth}{(a)}
          \fig{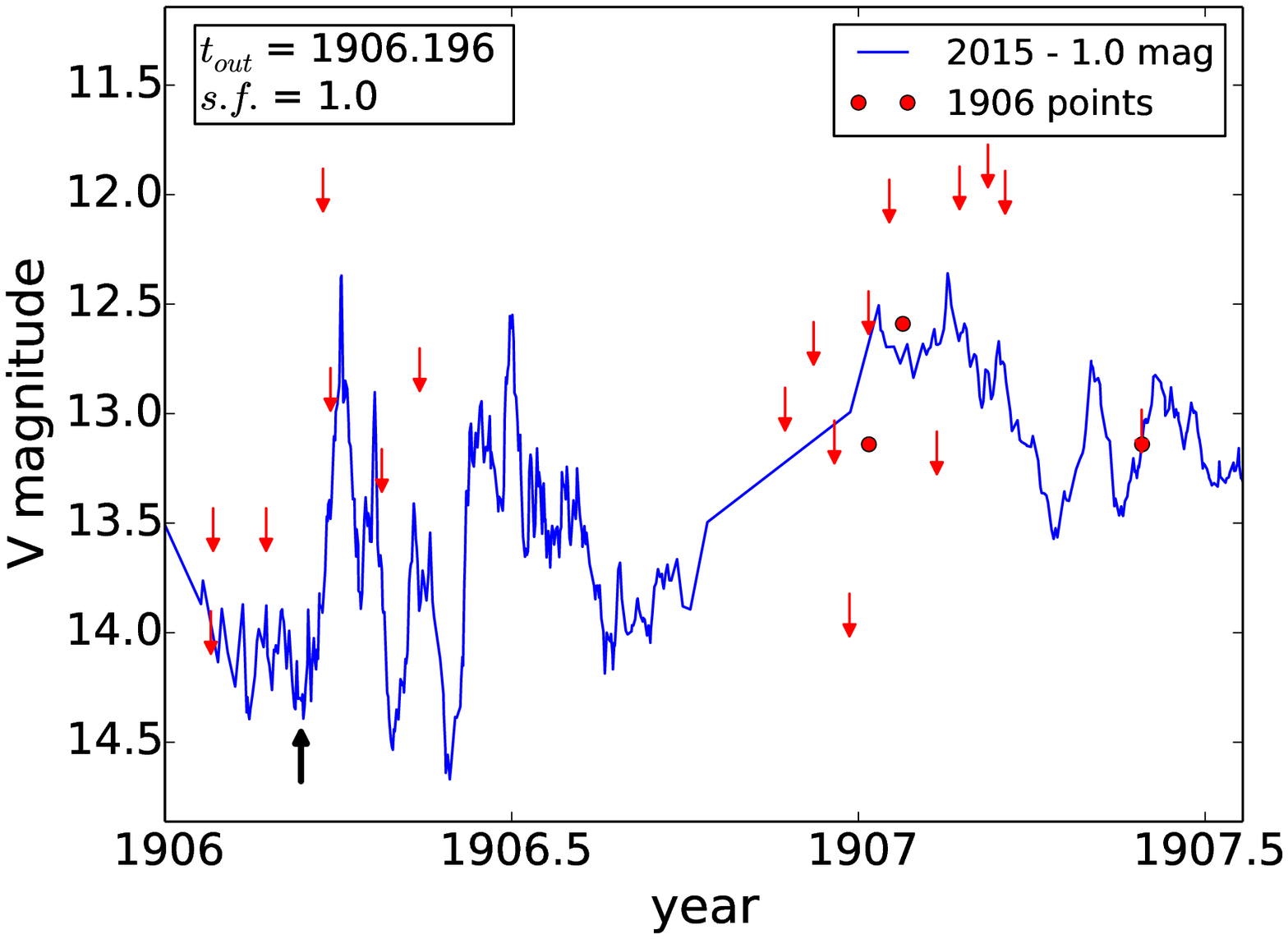}{0.33\textwidth}{(b)}
          }
\gridline{\fig{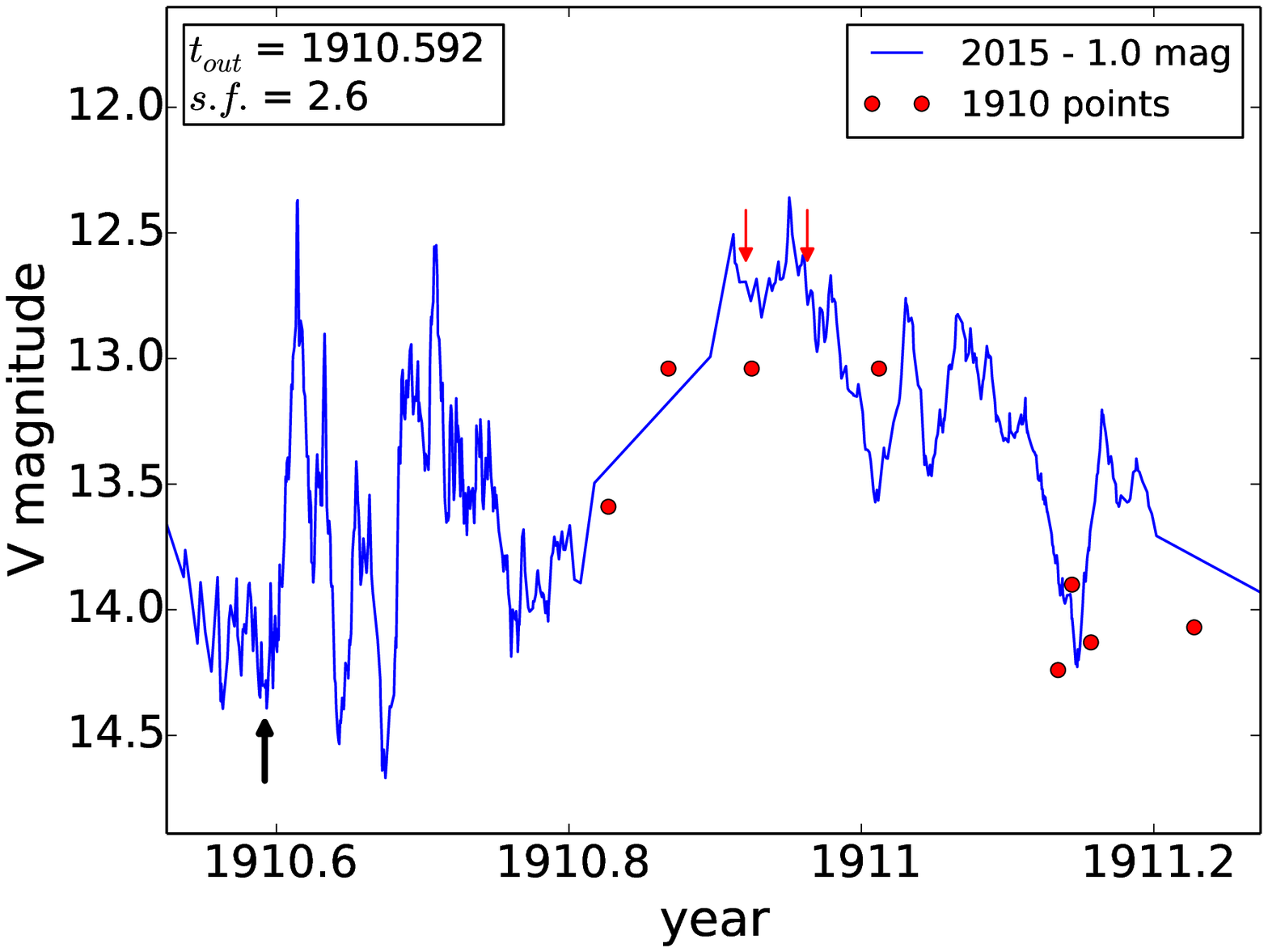}{0.33\textwidth}{(c)}
		  \fig{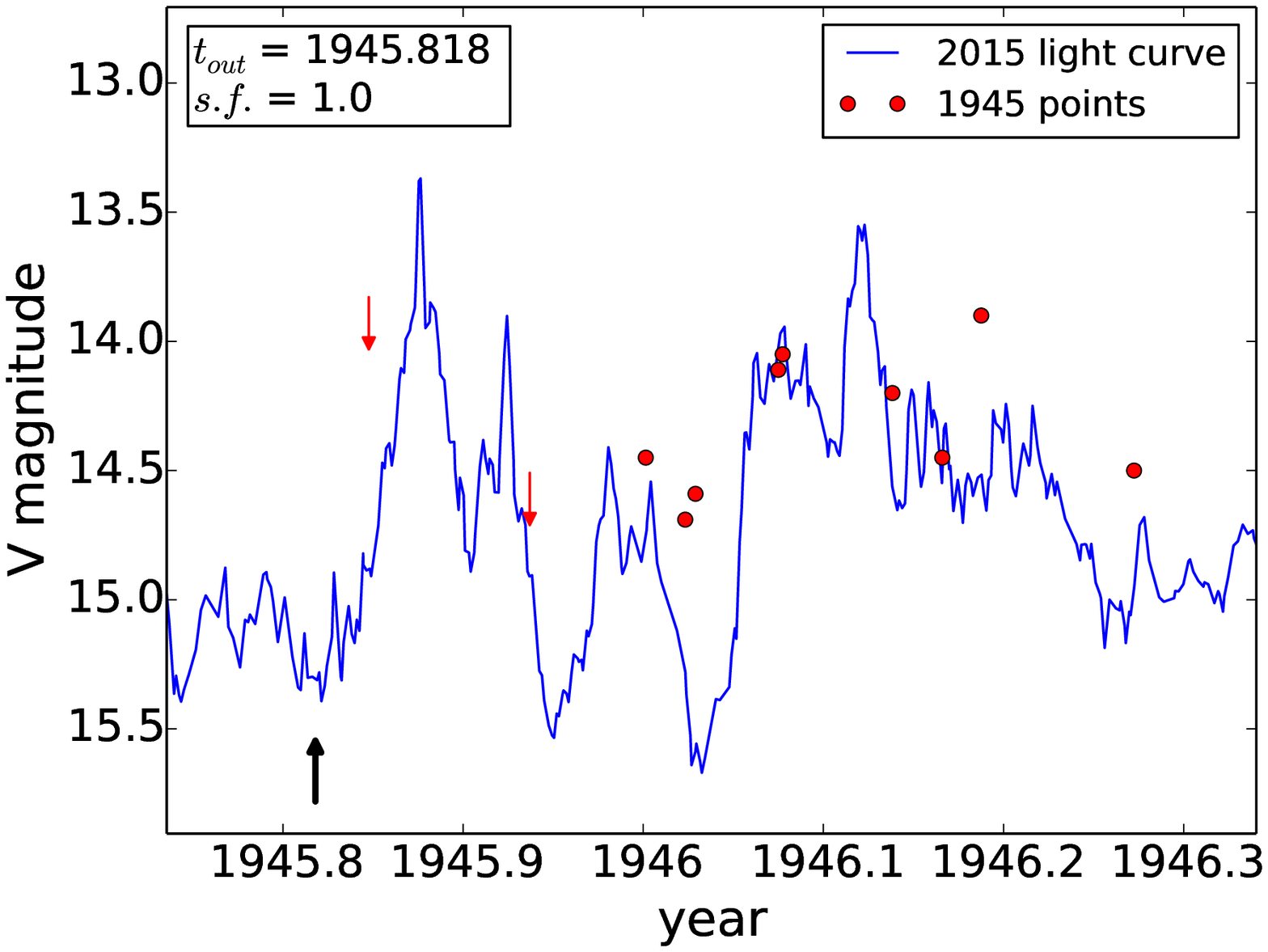}{0.33\textwidth}{(d)}
          }
\caption{Light-curve comparisons of rather poorly covered outbursts. Tips of the red thin arrows (pointing downward) indicate upper limits, whereas the black thick arrows (pointing upward) indicate the starting time of outbursts in the diagram. For the 1898 outburst (panel (a)), we have used the 2007 outburst light curve as a template because the 2015 light curve is not sufficiently long for this comparison. For the 1906 and 1945 outbursts (panels (b) and (d), respectively), the upper limits provide good constraints on the outburst timings.\label{fig:old_lc_comp1}}
\end{figure*}

\begin{figure*}
\gridline{\fig{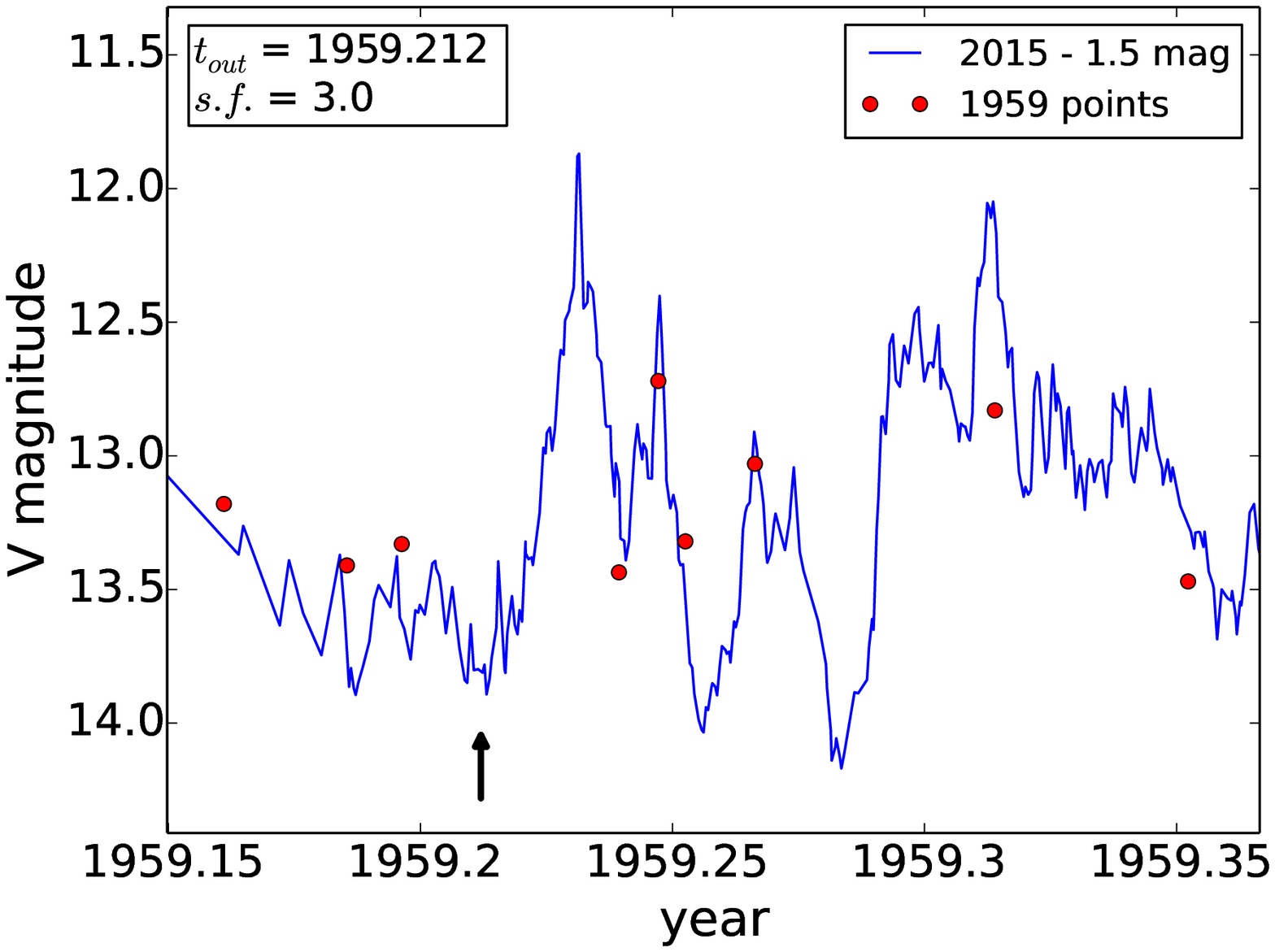}{0.33\textwidth}{(a)}
		  \fig{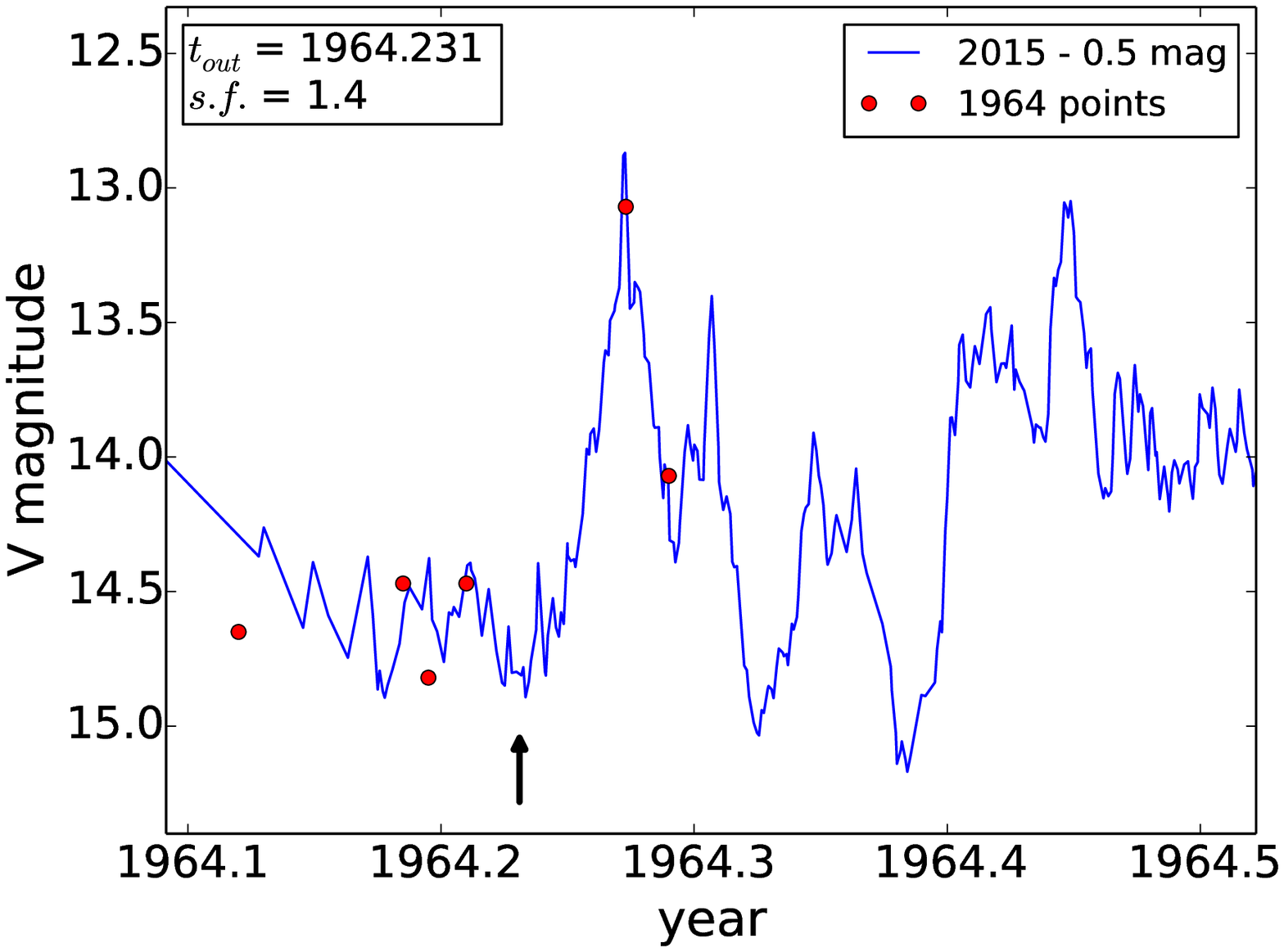}{0.33\textwidth}{(b)}
          }
\gridline{\fig{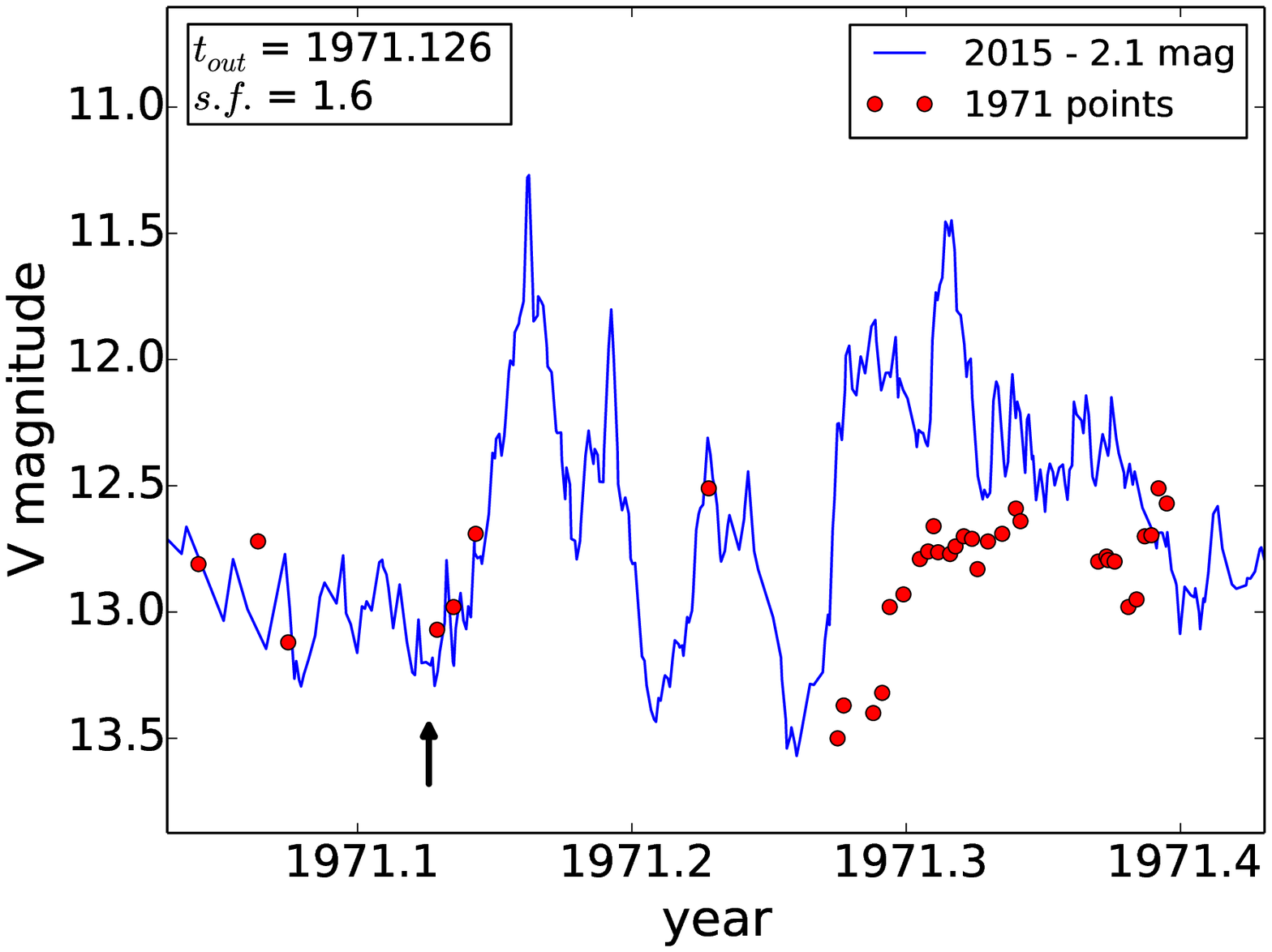}{0.33\textwidth}{(c)}
		  \fig{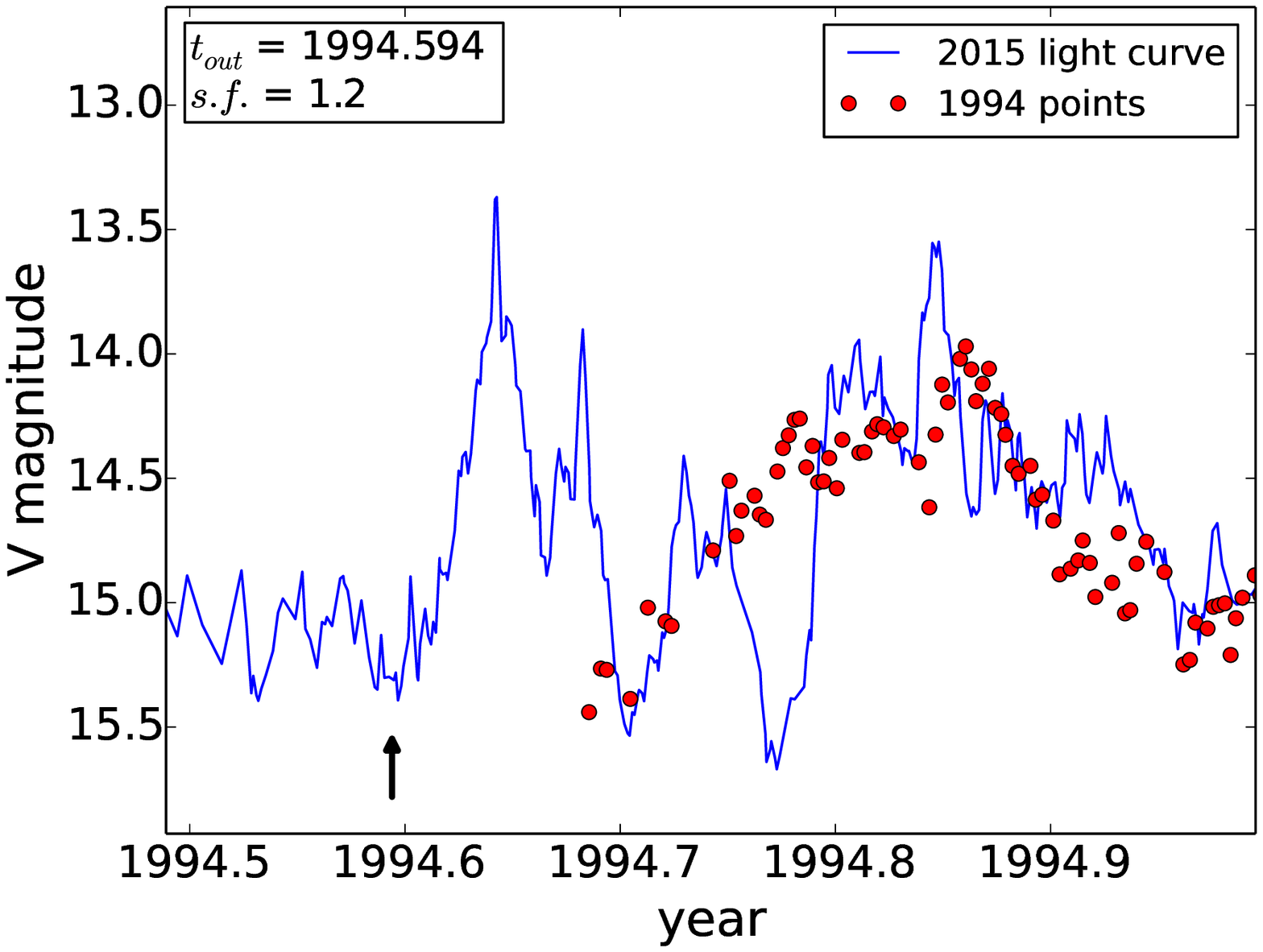}{0.33\textwidth}{(d)}
          }
\caption{Light-curve comparisons of rather poorly covered outbursts. The black thick arrows (pointing upward) indicate the starting time of outbursts in the diagram. Panel (c) shows that for the 1964 outburst, though we cannot determine the starting time of outburst, the high-brightness points indicate that the outburst has taken place at that time. For the 1959 and 1994 outbursts (panels (a) and (d), respectively), we missed the primary peaks, but the secondary and subsequent peaks are in agreement with the template light curve. \label{fig:old_lc_comp2}}
\end{figure*}

\begin{figure*}
\gridline{\fig{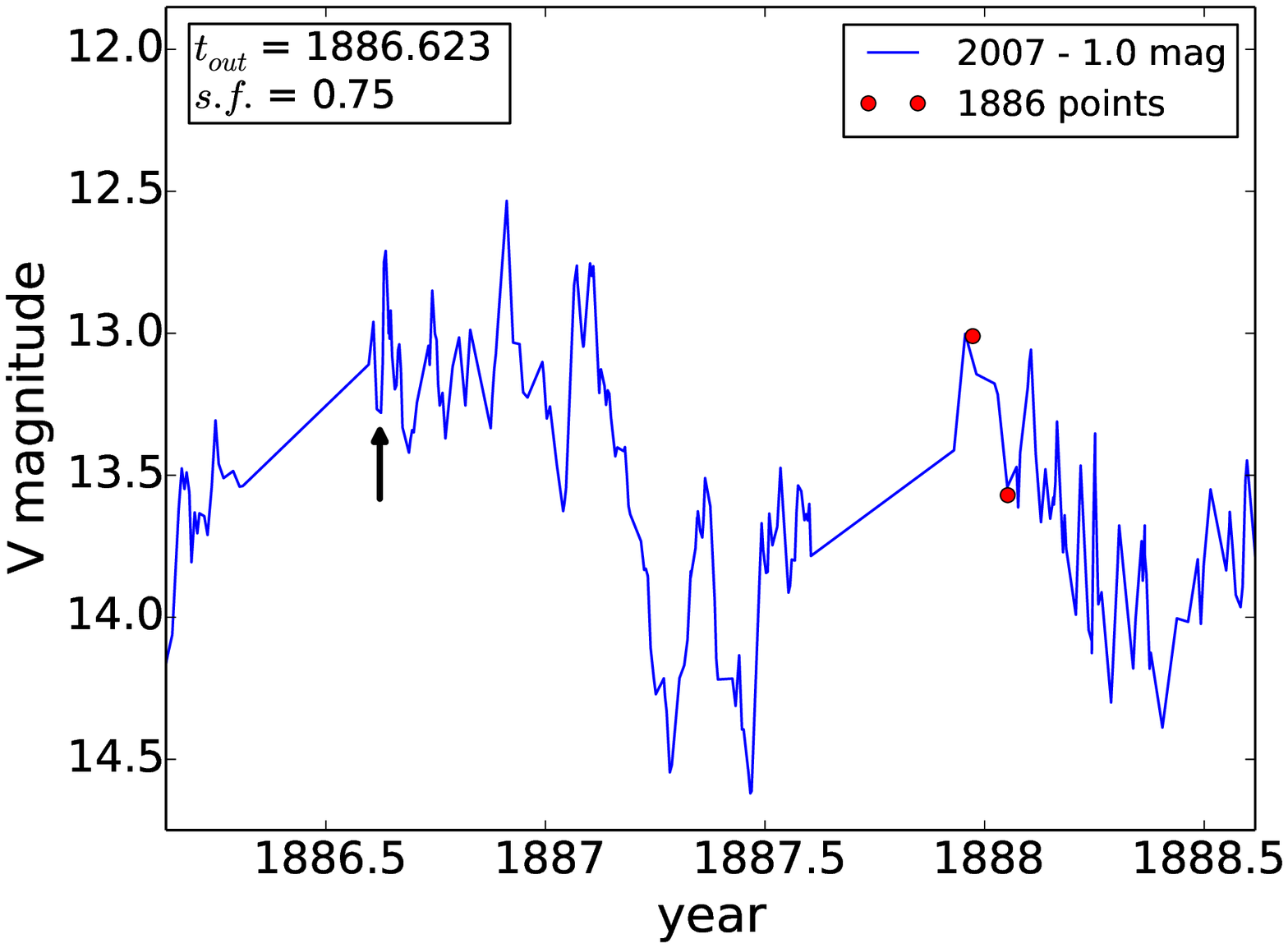}{0.33\textwidth}{(a)}
          \fig{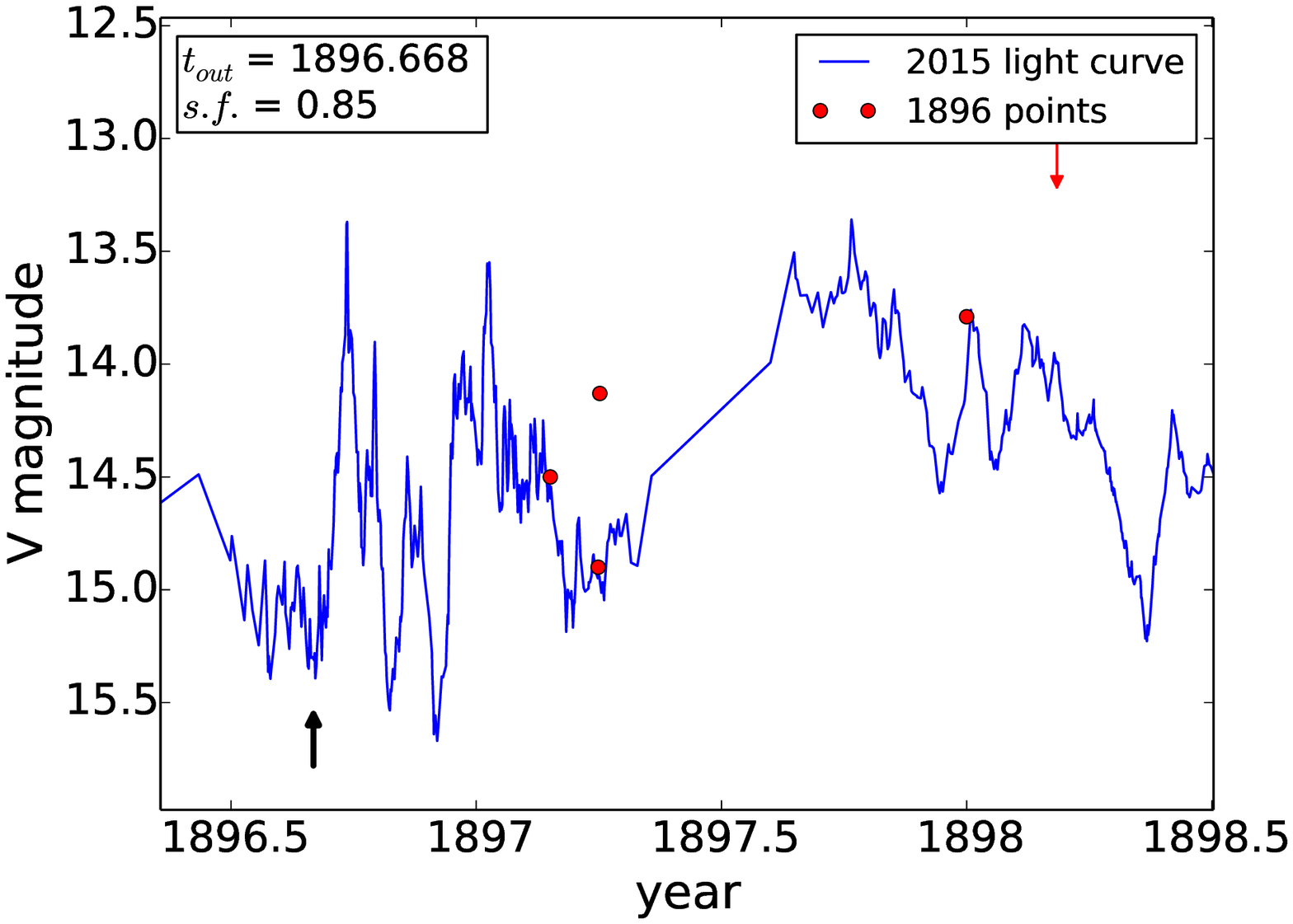}{0.33\textwidth}{(b)}
          }
\gridline{\fig{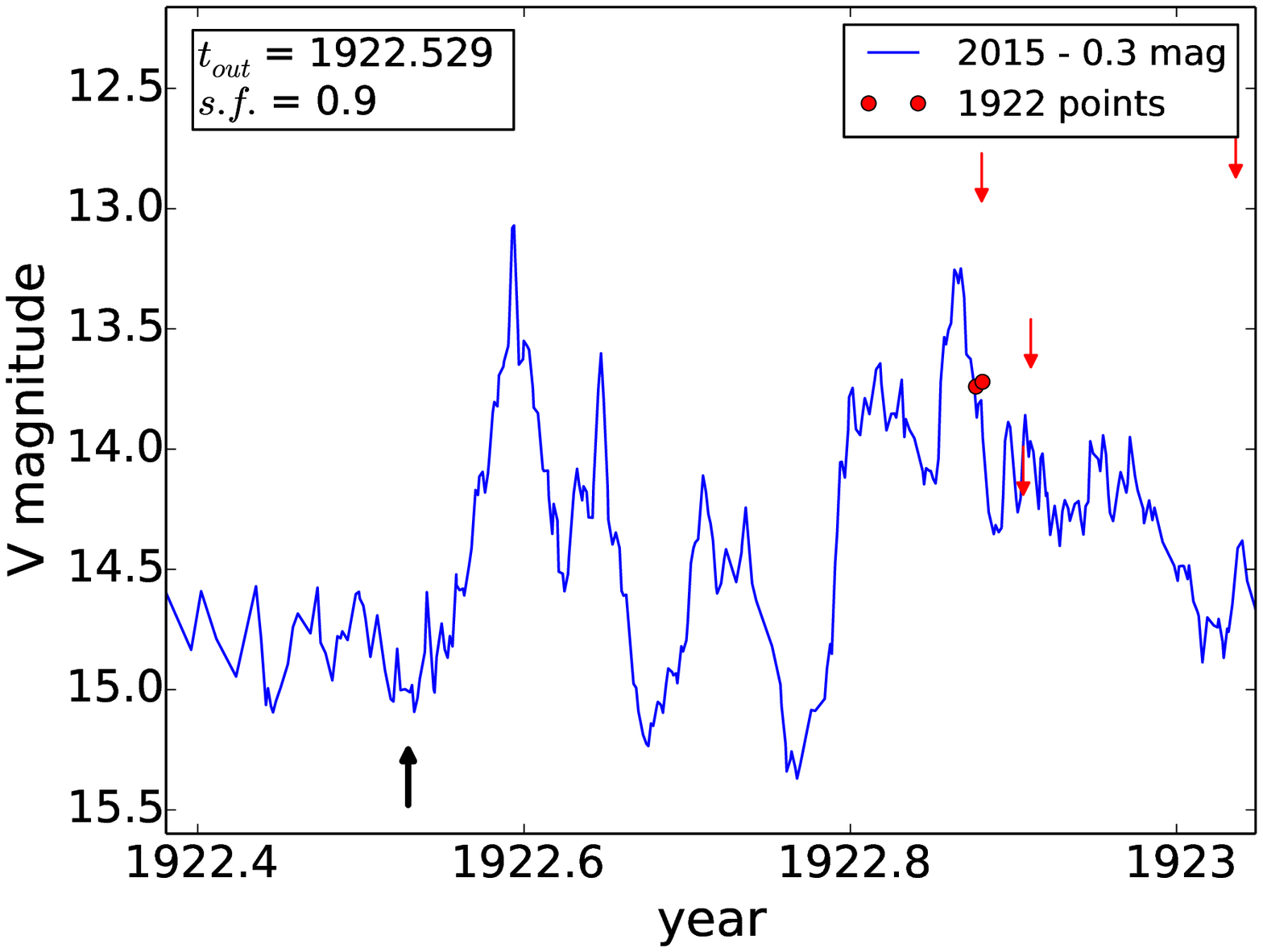}{0.33\textwidth}{(c)}
		  \fig{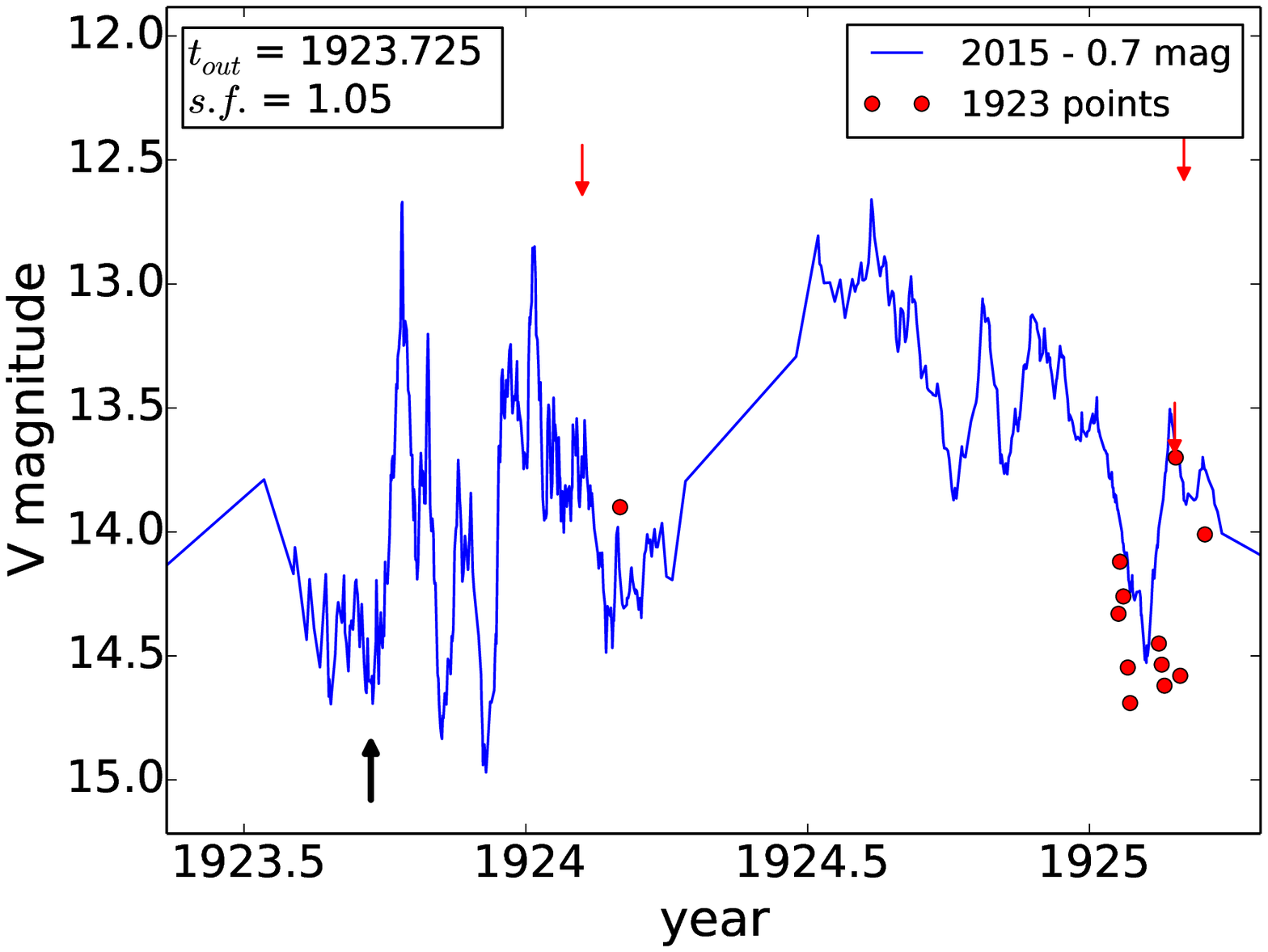}{0.33\textwidth}{(d)}
          }
\gridline{\fig{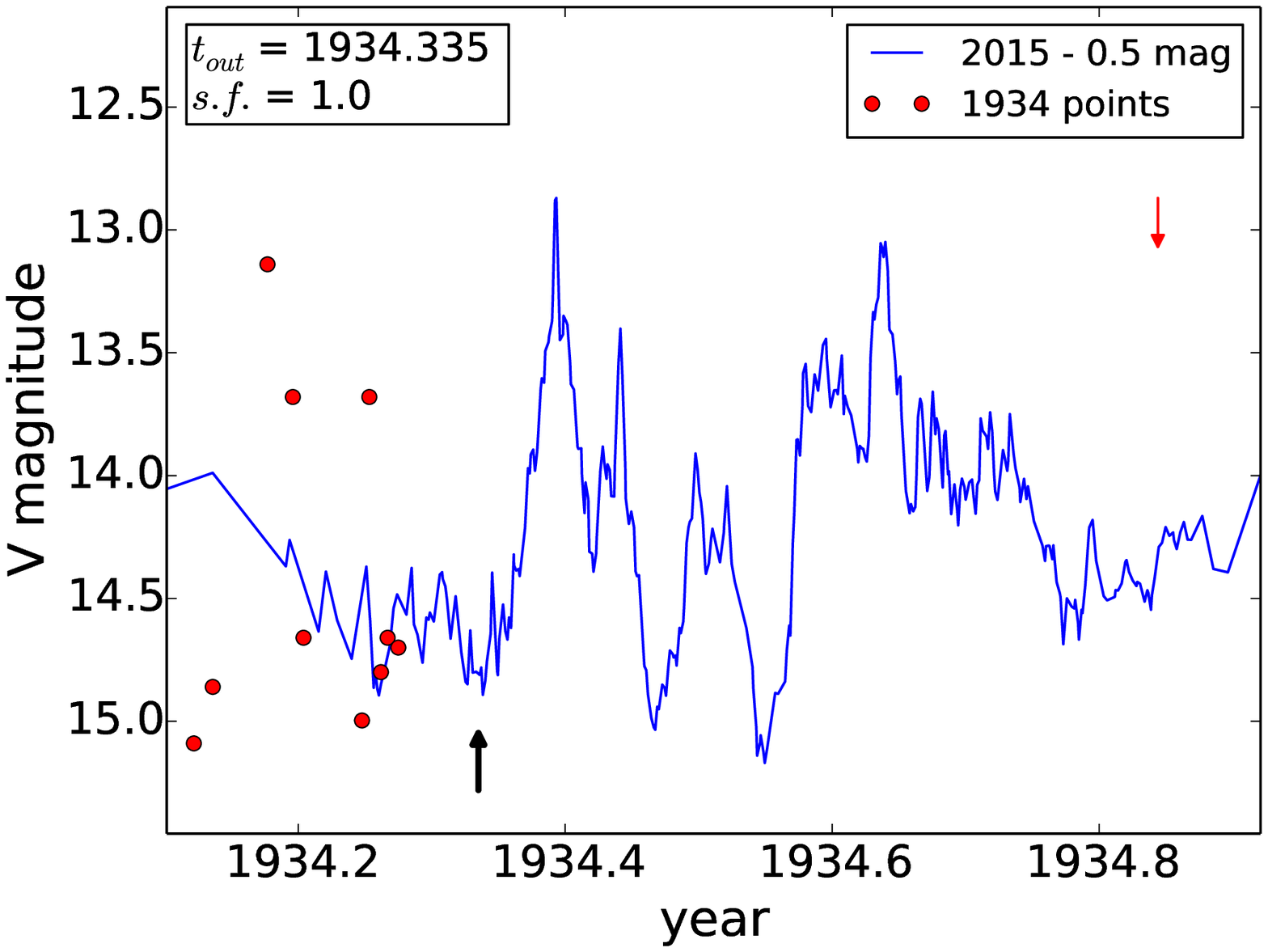}{0.33\textwidth}{(e)}
		  \fig{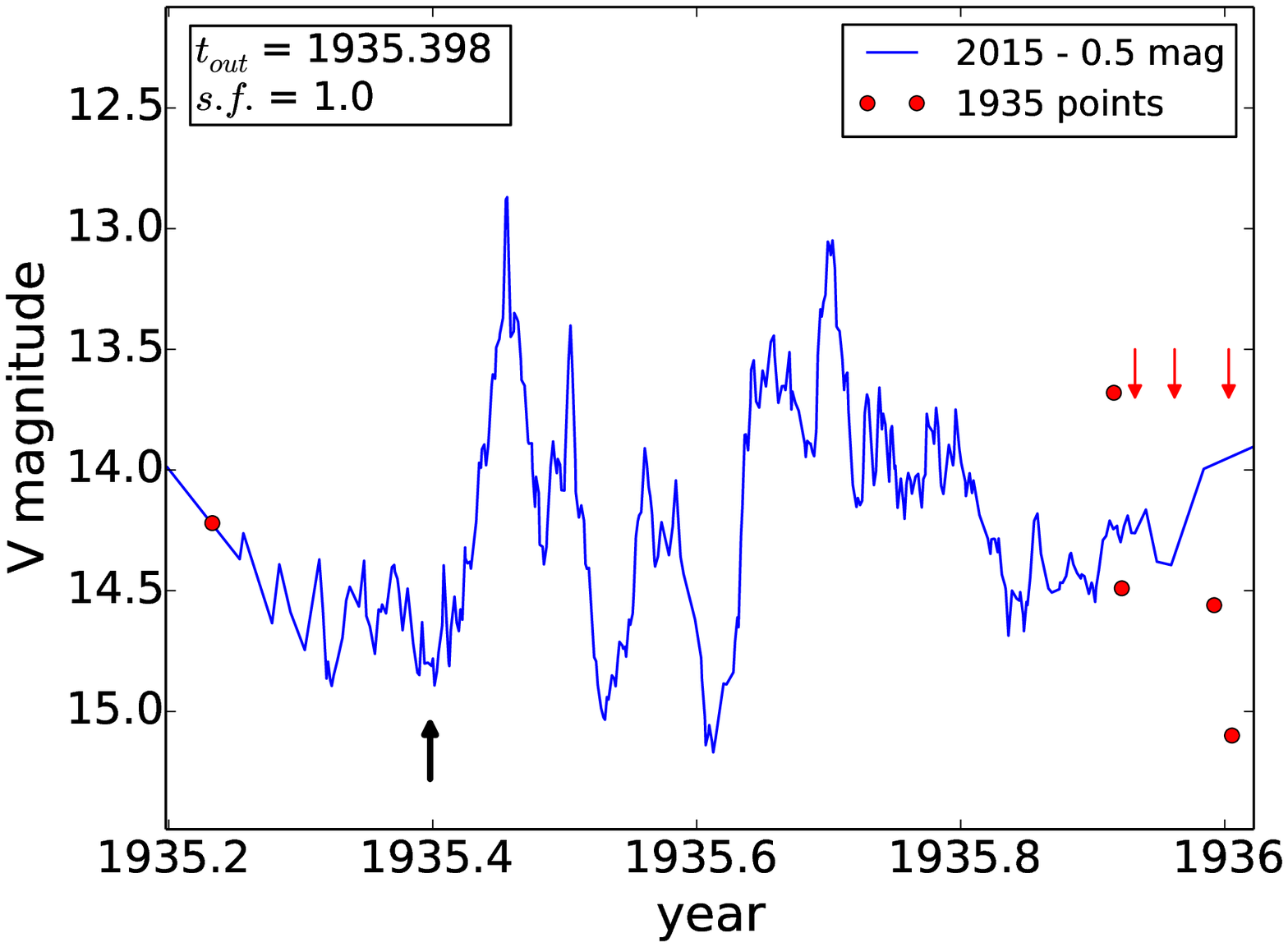}{0.33\textwidth}{(f)}
          }
\caption{Light-curve comparisons of very poorly covered outbursts. Tips of the red thin arrows (pointing downward) indicate upper limits, whereas the black thick arrows (pointing upward) indicate the starting time of outbursts in the diagram. For the 1886 outburst (panel (a)), we have used the 2007 light curve as the template because the 2015 light curve is not sufficiently long to make the comparison. The data points for these outbursts are too far away from the primary peak to make any meaningful timing estimates. \label{fig:old_lc_comp3}}
\end{figure*}

%\fi
This subsection mostly deals with the extended historical datasets on OJ~287 near impact flare epochs, predicted from our BBH central engine model. 
The availability of optical datasets on  OJ~287 extends to the late 19th century because of the fact that OJ~287 lies close to the ecliptic, and consequently was often unintentionally photographed in the past. New historical data points have been found by searching photographic plate archives for images containing OJ~287. The main dataset used in this work was compiled by Milan Basta and one of the authors (H.L., http://altamira.asu.cas.cz/iblwg/data/oj287) in 2006, using the previous compilations by other authors (A.S., L.O.T., T.P.). For the last 12 years our main database has been compiled by K.N. and S.Z. One of us (P.P.) made a light-curve compilation in 2012 including much of the data up to that point in time \citep{pih13,val13}. For the historical part, there has been a huge increase of data from the photographic plates of the Harvard plate collection, studied by R.H. He was able to evaluate numerous (almost 600) additional HCO plates not used before, partly because the object brightness was close to the plate magnitude limit, and he provided 364 additional measurements and 209 upper limits. Some earlier historical data were published by \cite{hud13} (HCO data) and by \cite{hud01} (Sonneberg Observatory data). 

We display in Figure~\ref{fig:lightcurve} the summary of the present state of available observational optical data points on OJ~287. 
The figure includes unpublished photographic data measurements obtained by R.H. from various astronomical photographic archives. The collection of the photometric data for the OJ~287/2015 campaign is described by \cite{val16}. S.Z. has collected the data and harmonized them in a uniform system.

As noted earlier, we are mostly interested in observational datasets around a number of impact flare epochs, predicted in our model.
Let us emphasize that many of these datasets are {\it not} employed while determining the parameters of our BBH central engine model for OJ~287. This  influenced us to find possible signatures of impact flares in the historical datasets on OJ~287.
For this purpose, we compare sections of observed data points on OJ~287 around the predicted impact flare epochs with a template light curve and search for the presence of possible patterns.
The  light curve from late 2015 to early 2017 is used as the standard template to compare with less-complete datasets from earlier major flare epochs that were created by secondary BH impacts according to our model. 
This is what we pursue in Figures~\ref{fig:correlation}, \ref{fig:correlation1}, \ref{fig:old_lc_comp1}, \ref{fig:old_lc_comp2}, and \ref{fig:old_lc_comp3}.
For smooth comparisons, we shift the 2015 outburst light curve, shown as line plots, backward in time by certain amounts such that the start times of outbursts coincide with epochs listed in Table~\ref{tab:orbit solution}.
In these figures, observational datasets are usually marked by points. For many epochs we only have upper limits for the brightness of OJ~287; these data points are marked by the tips of red thin arrows pointing downward. 

Influenced by \cite{val11a}, we introduce a certain ``speed factor" ($s.f.$) parameter which indicates how fast the earlier outburst has proceeded in comparison with the 2015 outburst. Termed as the $f$-parameter by \cite{val11a}, it depends on the velocity of the secondary BH and the distance of the impact site from the primary. 
It turned out that the rate at which the outburst takes place is about three times faster for impacts near the pericenter than for impacts at larger distance from the primary \citep{leh96}.
This is an important aspect, quantified by the dynamical timescale $t_{\rm dyn}$ of Table 3 in \cite{leh96}, while making detailed comparisons of our template flare light curve with actual observational data points.
We have also shifted the template light curve vertically by a certain magnitude (mentioned in the plot legends at the top-right corner of each plot) for matching the base levels of two outbursts. The variations in base levels arise because of long-term variations in the optical light curve as evident from Figure~\ref{fig:lightcurve}.

 We begin by displaying in Figures~\ref{fig:correlation} and \ref{fig:correlation1}  our comparisons of the 2015 template light curve with the well-documented outbursts of Table~\ref{tab:outburst}.
To quantify these comparisons, we compute Pearson correlation coefficients between the 2015 and the earlier outburst datasets, implemented using the {\it Correlation} routine of {\it  Mathematica} \citep{Mathematica}.
These coefficients are computed for restricted datasets that span two months and their values are listed below the panels. The bottom panels in Figure.~\ref{fig:correlation1} require further explanations. We do not compute the correlation coefficient for the 1947 outburst, as  we do not have sufficient data points to calculate it (this is despite the fact that the crucial sudden rise in brightness is well covered by observations during 1947). 
In panel (d) of Figure~\ref{fig:correlation1}, we display the expected light curve for the 2019 July outburst, and this is obtained by moving forward in time the 2007 light-curve data points. 
We invoked the dataset of the 2007 flare, as the predicted 2019 outburst is expected to be a periastron flare like the 2007 one in our BBH central engine model. An additional point worth noting is the low correlation coefficient of $0.65$ between the 2007 and 2015 outburst datasets; this is mainly due to the smaller rise in brightness of the first peak during the 2007 outburst as visible in panel (b) of  Figure~\ref{fig:correlation1}. 

In Figure~\ref{fig:old_lc_comp1} and Figure~\ref{fig:old_lc_comp2}, we compare a few less-well-covered outburst datasets with the 2015 light curve. { For the 1906 and 1945 outbursts, shown respectively in panels (b) and (d) of Figure~\ref{fig:old_lc_comp1}, the upper limits provide good constraints on outburst timings.
However, we required a shifted 2007 impact flare light curve to obtain a visual match with very few data points of the 1898 outburst, as shown in the panel (a) of Figure~\ref{fig:old_lc_comp1}. We employed the 2007 light curve as our standard outburst light curve; the 2015 dataset turned out to be not adequately long, as we compare datasets spanning roughly two years in panel (a) of Figure~\ref{fig:old_lc_comp1} (a more detailed study of the 1898 outburst is available in \cite{hud13}).

We now show comparisons for the 1959, 1964, and 1971 outburst datasets in panels (a), (b) and (c) of Figure~\ref{fig:old_lc_comp2}. Visual inspection reveals that impact flares most likely occurred at these predicted epochs. Unfortunately, we do not have archival data points for the primary peaks of these impact outbursts.
However, available data points from neighborhood epochs are consistent with our 2015 light curve. It is reasonable to infer from these figures that the available datasets are consistent with $17$ outburst epochs, and this is {\it seven} more than what is necessary to describe the BBH orbit.
Finally, in Figure~\ref{fig:old_lc_comp3}, we pursue comparisons of the remaining $6$ outbursts that are listed in Table~\ref{tab:orbit solution}.
Unfortunately, we have fairly poor observational data associated with the relevant epochs. 
Visual inspections suggest that the data points are not inconsistent with the model light curve. However, the available data points are too far from the relevant primary peaks to make any meaningful timing estimates.

A closer look at Figures~\ref{fig:correlation} and \ref{fig:correlation1} reveals that the outbursts with sufficient observed data points give high correlation coefficients with our template light curve. This clearly requires us to invoke the speed factor {\it s.f.}. The time evolution of the bubble of gas that is released as a result of the impact of the secondary on the accretion disk depends on the local disk conditions, the thickness of the disk, and the internal sound speed of the released bubble of gas; these quantities are encapsulated in the dynamical timescale or the speed factor {\it s.f.}. Once the speed factor is included, the listed high correlations suggest the possibility of predicting the general shape of the optical light curves associated with the future impact outbursts.

It should be noted that the major axis of the OJ~287 BBH eccentric orbit happens to lie in the disk plane during the disk crossing associated with the 2015 outburst  \citep{val16}.
This indicates that the BBH orbit is symmetric while going back and forward from its 2013 configuration.
Therefore, the 2019 impact is expected to occur in a manner similar to the 2007 impact with the same speed and at the same impact angle. This should result in an optical light curve which has a high resemblance to the well-documented 2007 impact flare light curve.
These are our main arguments behind  panel (d) of Figure~\ref{fig:correlation1}, where we are essentially predicting the shape of the light curve for the 2019 impact outburst. Extending these arguments, we may state that the 2022 impact event should resemble the 2005 impact outburst, while the 2031 impact light curve should resemble the one associated with the 1995 event in OJ~287, and so on. It applies also to the other future impact events, listed in Table~\ref{tab:orbit solution}. 
These considerations open up interesting possibilities during the next outburst, and this is 
tackled in the next subsection.

\subsection{ Testing the BH No-Hair Theorem with the Predicted 2019 Impact Flare}

This subsection revisits an idea that was explored in detail by some of us earlier \citep{val11a}:
testing a formulation of the BH no-hair theorem which requires that the dimensionless quadrupole moment of a BH ($q_2$) should fulfill the relation $q_2 = -q\, \chi^2 $ where $q=1$ in GR \citep{Thorne80}. For obtaining observationally relevant constraints on $q$, we invoke our GR-based estimate for $\gamma$, namely $\gamma =1.2917$, and treat the above $q$ as a free parameter while numerically determining the BBH orbit. 
Recall that the $q$ parameter enters the BBH dynamics via the classical spin-orbit coupling term $\ddot {\bf x}_{Q}$ in the equations of motion. 

\begin{figure*}
\centering
\gridline{\fig{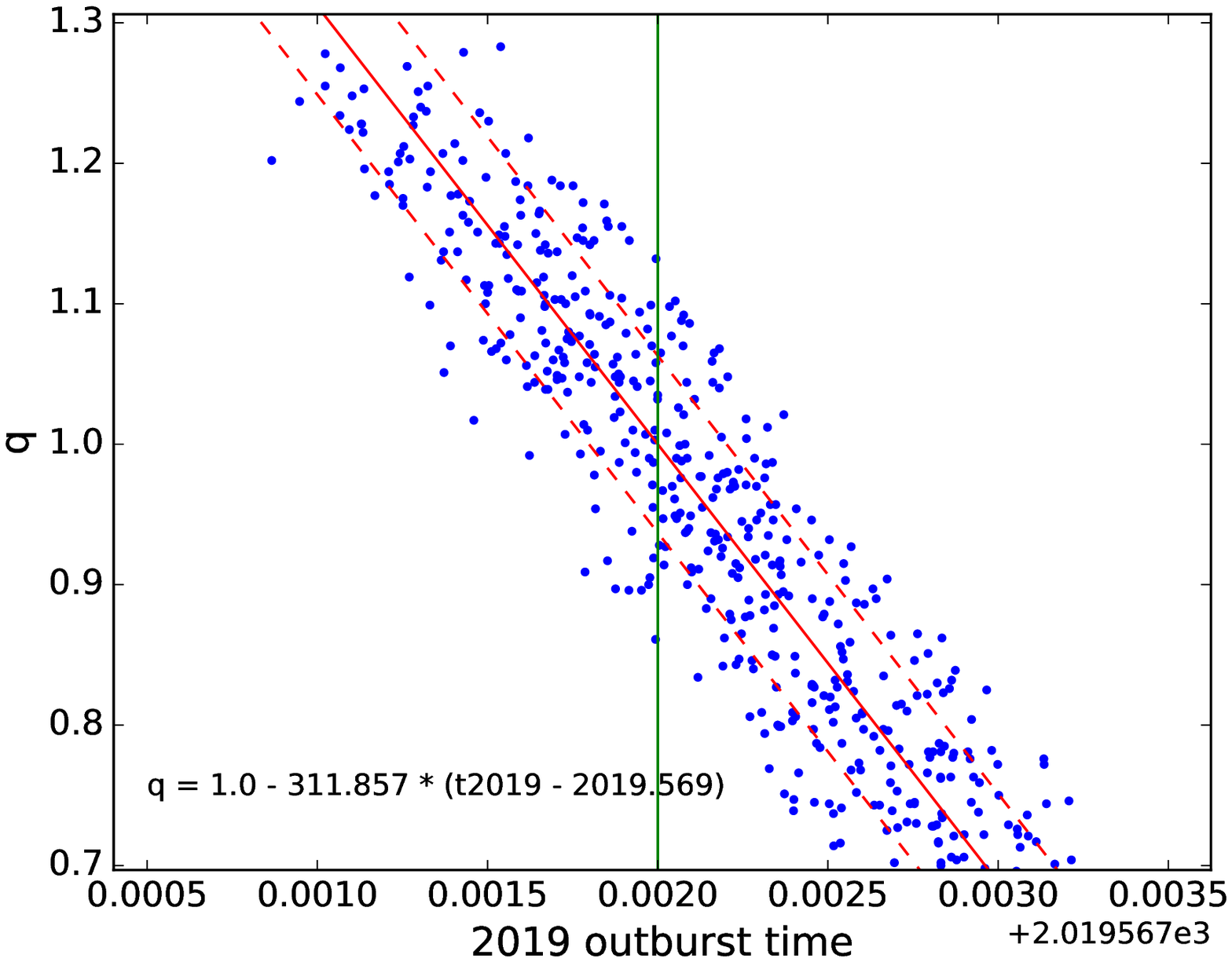}{0.45\textwidth}{(a)}
		  \fig{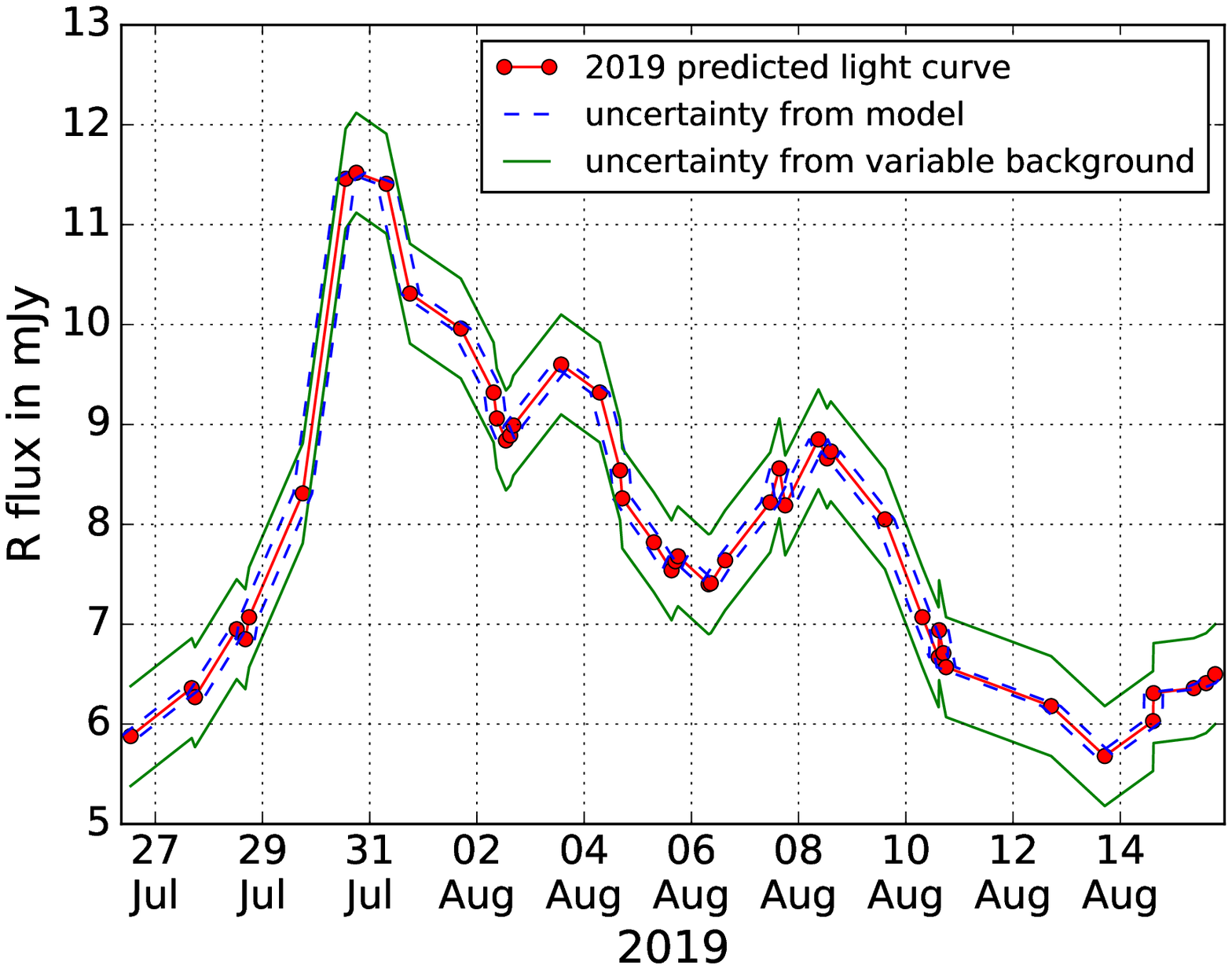}{0.45\textwidth}{(b)}
          }
\caption{
The left panel shows that an accurate timing of the predicted 2019 outburst should lead to a $10\%$-level test of the BH no-hair theorem by plotting the correlation between the starting time of the 2019 outburst and our $q$ parameter that appears in the $ q_2 = -q \, \chi^2$ relation for the primary BH in OJ~287. 
The solid red line shows our fit to the observed correlation and the dashed red lines indicate its $1\sigma$ deviation (the fit equation is given in the lower-left corner of the plot).
The green vertical line represents the expected outburst starting time from the model.
The right-hand panel shows the expected light curve for the 2019 July outburst, based on our ability to predict the shape and evolution of the impact flare light curves. The envelope lines outline the uncertainty arising from the variable background light level coming from the jet of OJ~287 (wider envelope), and the model uncertainly (inner envelope).
\label{fig:no-hair}}
\end{figure*}

It turns out that the 2019 outburst time is highly correlated with the $q$ parameter. We plot the correlation of this ``no-hair" parameter $q$ against the time of the rapid rise of the flare ($t_{2019}$) in the left panel of Figure~\ref{fig:no-hair}, where the slanted lines show the expected correlation and the $1\sigma$ deviation from it. 
For this numerical experiment, we employed the BBH parameters listed in Table~\ref{tab:parameters}. 
Additionally, we extracted a numerical fit that provides the accuracy with which we can estimate $q$ in terms of $t_{2019}$: 
\begin{eqnarray}
\label{eqn_qfit}
q &=& 1.0 - 311.857 (t_{2019} -2019.569)\,.
\end{eqnarray}
This formula clearly shows that an accurate $t_{2019}$ measurement should allow us to determine the $q$ parameter below the $10 \%$ level.
Indeed, this provides a substantial improvement over the present $q$ estimate that confines it {\it only } in the 0.5--1.5 range.
Currently, there exist no other observational constraints on $q$ even at this rather relaxed accuracy level.

We now turn our attention to the feasibility of observing the next predicted outburst. 
In our BBH central engine model, the next outburst is expected to peak on {\bf{July 31}}, 2019.
However, observing OJ~287 during the expected 2019 outburst window is practically impossible from the ground: the angular distance in the sky between the Sun and OJ~287 is only $\sim 6^\circ$ at the beginning of this event, and it goes down to $4^\circ$ by the time of peak brightness.
Unfortunately, objects at small angular distances from the Sun are difficult to observe even from a satellite in Earth's orbit, owing to the high background caused by intense sunlight.

In any case, it will be useful to have good light curves for the 2018--2019 and 2019--2020 observing seasons. We provide the expected light curve of OJ~287 that spans a few weeks around July 31, 2019 in panel (b) of Figure~\ref{fig:no-hair}.
Especially important would be to determine the epoch of rapid flux rise associated with the primary peak of the 2019 impact flare. However, even if the primary peak is missed, the secondary peak can still provide some information on the flare timing. This was the case in observations of the 1994 outburst. 
We infer from panel (d) of Figure~\ref{fig:old_lc_comp2} that the primary peak of OJ~287 was missed owing to the closeness of OJ~287 to the Sun during the 1994 outburst.
However, the visual correlation of the secondary flares with our 2015 template light curve allows us to state that the starting time of the outburst was around 1994.59, and this is consistent with our BBH model. Similarly, it will be useful to organize an observational campaign to observe the secondary and subsequent peaks using ground-based facilities around the 2019 impact flare epoch.

The flare of 2019 July 31 would be best observed from a satellite observatory which is far from Earth, such as the STEREO-A solar telescope or the {\it Spitzer Space Telescope}. As an observational target OJ~287 is relatively easy, as it brightens to $\sim 13$ mag in the optical $R$ band. The only problem is to find a space telescope that is able to do optical photometry during the rapid rise of flux, on July 29--31, 2019!

\section{Conclusions and Discussion} \label{sec:dis_con}

The present paper provides the most up-to-date and improved description for the binary black hole central engine of OJ~287.
This is mainly because of the use of an improved PN prescription to describe the BBH orbit evolution.
We incorporate in the BBH dynamics the effects of next-to-next-to-next-to leading (or quadrupolar) order GW emission. This includes effects due to the dominant-order hereditary contribution to the GW-induced inspiral. 
It turns out to be crucial to incorporate the effect of hereditary contributions to GW emission on the BBH dynamics, and we develop an approach to model the effect into ${\ddot{\vek x}}$ with the help of an unknown parameter $\gamma$.
The observationally determined value $\gamma_{obs}$ shows remarkable agreement with its GR-based estimate $\gamma_{GR}$, obtained by adapting GW phasing formalism for eccentric binaries. 
This formalism is required to construct accurate inspiral templates to model GWs from compact binaries that are inspiraling along PN-accurate eccentric orbits. 
Furthermore, we incorporate  next-to-leading-order spin-orbit contributions to the compact binary dynamics, influenced by  \cite{BBF_06} and \cite{wil17}.
This leads to a noticeably different estimate for the Kerr parameter of the primary BH, namely $\chi= 0.381 \pm 0.004$, compared to $\chi=0.313$ in \cite{val16}. 
Additionally, the rate of decay of the binary orbit is slower than in earlier models by about $6.5\%$ \citep{val10b}.
The improved description allows us to demonstrate excellent agreement between the observed impact flare timings and those predicted from the BBH central engine model.

These improvements should allow us to employ the BBH central engine model to test GR in the strong-field regime that at present is not accessible to any other observatories or systems. 
The first such test is possible in 2019 July. The next major flare will peak on July 31, around noon GMT in our model. 
The model without higher-order gravitational radiation reaction terms gives the brightness peak 1.57 days earlier, in the early hours of July 30 GMT. These two models are easily differentiated by observations, provided we are able to monitor OJ~287 during late July 2019. 
The closeness to the Sun in the sky makes such an effort extremely difficult.
However, a successful observational campaign should provide us the unique opportunity 
to test the black hole no-hair theorem at the $\sim 10 \%$ level during the present decade.
Additionally, we demonstrate the possibility of predicting the general shape of the expected optical light curve of OJ~287 during the impact flare season. 
This should be helpful in analyzing the optical light curve of OJ~287 during the next two accretion impact flares, expected to happen during 2019 and 2022. 
These observational campaigns will be challenging owing to the apparent closeness of the blazar to the Sun.
However, the monitoring of these impact flares should allow us to test general relativity in the strong-field regime that is characterized by $ (v/c) \approx 0.25$ and $ m \approx 18\times 10^9 M_{\odot}$. 
It will be exciting to extend the preliminary results, displayed in Figure 6 of \cite{val12}, that provided an independent estimate for the mass of the central BH in OJ 287. It turned out that the dynamically estimated total mass in OJ~287 and the measured absolute magnitude of the bulge of the  host galaxy is fully consistent with the black hole mass - K-magnitude correlation pointed out in \cite{kormendy11}.

\acknowledgments
L.D. acknowledges the hospitality of Tuorla Observatory, University of Turku, where part of this work was carried out. S.Z. acknowledges support from NCN grants 2012/04/A/ST9/00083 and 2018/09/B/ST9/02004. RH acknowledges GACR grant 13-33324S. P.P. acknowledges support from the Academy of Finland, grant 274931. A.V.F. has been supported by the Christopher R. Redlich Fund, the TABASGO Foundation, and the Miller Institute for Basic Research in Science (UC Berkeley). His work was conducted in part at the Aspen Center for Physics, which is supported by National Science Foundation (NSF) grant PHY-1607611; he thanks the Center for its hospitality during the supermassive black holes workshop in June and July 2018.

%% Appendix material should be preceded with a single \appendix command.
%% There should be a \section command for each appendix. Mark appendix
%% subsections with the same markup you use in the main body of the paper.

%% Each Appendix (indicated with \section) will be lettered A, B, C, etc.
%% The equation counter will reset when it encounters the \appendix
%% command and will number appendix equations (A1), (A2), etc. The
%% Figure and Table counter will not reset.

\appendix

\section{PN-Accurate Expressions for the Secular Evolution of the BBH's Orbital Phase}
\label{app:kdndldetdl}

The 3PN-accurate expression for $k$ (fractional rate of advance of periastron), extracted from \cite{KG06}, reads
\begin{eqnarray}
\label{eqn:k_3PNl}
k & 
= \frac{\displaystyle 3 \,\xi}{\displaystyle 1 - e_t^2 }  
+ \frac{\displaystyle \xi ^{2} }{\displaystyle 4 ( 1 - e_t^2 )^2 } 
\biggl \{ 78 - 28 \eta + ( 51 - 26 \eta ) e_t^2 \biggr \}
+ \frac{\displaystyle \xi^3 }{\displaystyle 128 ( 1 - e_t^2 )^3 }
\Big \{ 18240 - 25376 \eta 
\nonumber \\&
+ 492 \pi^2 \eta + 896 \eta ^2 
+ ( 28128 - 27840 \eta + 123 \pi^2 \eta + 5120 \eta^2 ) e_t^2
+ ( 2496 - 1760 \eta 
\nonumber \\&
+ 1040 \eta^2  ) e_t^4
+ \bigl [ 1920 - 768 \eta + ( 3840 - 1536 \eta ) e_t^2 \bigr ] \sqrt{1 - e_t^2}
\Big \}\,,
\end{eqnarray}
where $\xi = (G\,m\,n/c^3)^{2/3}$ such that $n =2\, \pi/P_b$ is the mean motion and $e_t$ provides the eccentricity parameter that enters the PN-accurate Kepler equation of \cite{MGS}. The mean motion ($n$) and the eccentricity ($e_t$) of the system evolve with time due to emission of GWs.

Following \cite{BDI}, we write fully 2PN-accurate expressions for the temporal evolutions of $n$ and $e_t$ owing to GW emission in terms of certain ``instantaneous" and ``tail" contributions as 
\begin{subequations}
\begin{eqnarray}
\left( \frac{ d n }{ dl } \right)^{\rm {2PN}} &=& \left( \frac{ d n }{ dl } \right)^{\rm Inst} + \left( \frac{ d n }{ dl } \right)^{\rm Tail}\,,\\
\left( \frac{ d e_t }{ dl } \right)^{\rm 2PN} &=& \left( \frac{ d e_t }{ dl } \right)^{\rm Inst} + \left( \frac{ d e_t }{ dl } \right)^{\rm Tail}\,,
\end{eqnarray}
\end{subequations}
where we used $ n\, dt = dl $ to obtain $(dn/dl,de_t/dl)$ expressions from their $(dn/dt,de_t/dt)$ counterparts available in the literature.
The explicit expressions for these instantaneous contributions, appearing at the Newtonian, 1PN, and 2PN reactive orders, which depend only on the state of the binary at the usual retarded instant, are available in \cite{KG06}. 
The relevant differential equation for $n$ reads
\begin{eqnarray}
\left (\frac{ d n }{ dl } \right )^{\rm Inst} & = 
{\xi}^{5/2}\, n \, \eta
\biggl \{
%\dot{n}^{\rm N}
\frac{\displaystyle 1}{\displaystyle 5 (1 - e_t^2)^{7/2} }
\Biggl [ 96 + 292 e_t^2 + 37 e_t^4 \Biggr ]
%\nonumber \\&
+
%+ \dot{n}^{\rm 1PN}        
%\dot{n}^{\rm 1PN} & =
\frac{\displaystyle {\xi} }{\displaystyle 280 (1 - e_t^2)^{9/2} }
\Biggl [ 
20368 - 14784 \eta 
\nonumber \\&
+ ( 219880 - 159600 \eta ) e_t^2
%\nonumber \\&
+ ( 197022 - 141708 \eta ) e_t^4
+ ( 11717 - 8288 \eta ) e_t^6
\Biggr ] 
\nonumber \\&
+
%+ \dot{n}^{\rm 2PN}
%\dot{n}^{\rm 2PN} & =
\frac{\displaystyle {\xi}^{2} }{\displaystyle 30240 (1 - e_t^2)^{11/2} }
\Biggl [ 
12592864 - 13677408 \eta + 1903104 \eta^2
%\nonumber \\&
+ ( 133049696 - 185538528 \eta 
\nonumber \\&
+ 61282032 \eta^2 ) e_t^2
+ ( 284496744 - 411892776 \eta + 166506060 \eta^2 ) e_t^4
+ ( 112598442 
\nonumber \\&
- 142089066 \eta 
+ 64828848 \eta^2 ) e_t^6
+ ( 3523113 - 3259980 \eta + 1964256 \eta^2 ) e_t^8
\nonumber \\&
+ 3024
( 96 + 4268 e_t^2 + 4386 e_t^4 + 175 e_t^6 )
 ( 5 - 2 \eta ) \sqrt{1 - e_t^2}
\Biggr ] 
\biggr \}\,.
\end{eqnarray}
It is important to note that the $e_t$ contributions are exact while restricting our attention to the 
instantaneous contributions. The associated differential equation for $e_t$ reads
\begin{eqnarray}
\left ( \frac{ d {e}_t }{ dl } \right )^{\rm Inst}& =
- {{\xi}}^{5/2} \, \eta \, e_t
\biggl \{
%\dot{e}_t^{\rm N}
%\dot{e}_t^{\rm N} & =
\frac{\displaystyle 1 }{\displaystyle 15 (1 - e_t^2)^{5/2} }
\Biggl [ 304 + 121 e_t^2 \Biggr ]
\,,
+
% \dot{e}_t^{\rm 1PN}      
%\dot{e}_t^{\rm 1PN} & =
\frac{\displaystyle {\xi} }{\displaystyle 2520 (1 - e_t^2)^{7/2} }
\Biggl [ 
340968 - 228704 \eta
\nonumber \\&
+ ( 880632 - 651252 \eta ) e_t^2
+ ( 125361 - 93184 \eta ) e_t^4
\Biggr ] 
+
% \dot{e}_t^{\rm 2PN}
%\dot{e}_t^{\rm 2PN} & =
\frac{\displaystyle {\xi}^{2} }{\displaystyle 30240 (1 - e_t^2)^{9/2} }
\Biggl [ 
20815216
\nonumber \\&
- 25375248 \eta + 4548096 \eta^2
+ ( 87568332 - 128909916 \eta + 48711348 \eta^2 ) e_t^2
\nonumber
\\
& \quad
+ ( 69916862 - 93522570 \eta + 42810096 \eta^2 ) e_t^4
+ ( 3786543 - 4344852 \eta 
\nonumber \\&
+ 2758560 \eta^2 ) e_t^6
+ 1008 
( 2672 + 6963 e_t^2 + 565 e_t^4 )
( 5 - 2 \eta ) \sqrt{1 - e_t^2}
\Biggr ] 
\biggr \}
\,.
\end{eqnarray}

 We adapt an approach, detailed in Section~III of \cite{THG}, to incorporate 1.5PN-order tail contributions to 
 the differential equations for $n$ and $e_t$ in an accurate and efficient manner.
 These  dominant-order tail contributions 
 arise from the nonlinear interactions between the quadrupolar 
 gravitational radiation field and the mass monopole of the source and are 
 nonlocal in time. It is convenient to define these 1.5PN-order contributions to 
 far-zone energy and angular momentum fluxes, and therefore to 
 the differential equations for $n$ and $e_t$ 
 in terms of certain 
 eccentricity enhancement functions $\varphi(e_t)$ and $\tilde{\varphi}(e_t)$ \citep{BS93,RS97}.
 %http://adsabs.harvard.edu/abs/1993CQGra..10.2699B
 %http://adsabs.harvard.edu/abs/1997CQGra..14.2357R
 Tail contributions to the differential equations for $n$ and $e_t$ in terms of these enhancement functions read
 \begin{eqnarray}
\left ( \frac{d\,n}{dl} \right )^{\rm Tail} &=& \frac{\displaystyle 384}{\displaystyle 5}\, \frac{\displaystyle c^3}{\displaystyle G\,m}\, \eta \, \pi\, \xi^{11/2}\,\varphi(e_t)\,, \\
 \left ( \frac{d\,e_t}{dl} \right )^{\rm Tail}  &=& -\frac{394}{3}\, \eta \, \pi\,\xi^4\, 
 \biggl \{  \frac{192}{985\, e_t}\,{\sqrt{1 - e_t^2}}\,
 \left [\sqrt{1 - e_t^2}\,\varphi(e_t) - \tilde{\varphi}(e_t)\right ] \biggr \}\,.
\end{eqnarray}
 
 Owing to the hereditary nature of tail effects, 
 these functions are usually given 
in terms of infinite sums of Bessel functions  $J_n(n e_t)$ and their   derivatives with respect to $(n\,e_t)$.
We invoke a technique, detailed by \cite{THG}, which allows us to express these 
enhancement functions in terms  certain rational functions 
of $e_t$, and the final results are
\begin{eqnarray}
 \varphi(e_t) &=  \biggl(1 + 7.260831042\,e_t^2 + 5.844370473\,e_t^4 + 0.8452020270\,e_t^6 
 + 0.07580633432\,e_t^8 
 \nonumber \\&
 + 0.002034045037\,e_t^{10}\biggr)
 \biggl /
 \biggl(1 - 4.900627291\,e_t^2 + 9.512155497\,e_t^4 
 \nonumber \\&
 - 9.051368575\,e_t^6 
 + 4.096465525\,e_t^8 - 0.5933309609\,e_t^{10} 
 - 0.05427399445\,e_t^{12} 
 \nonumber \\ &
 - 0.009020225634\,e_t^{14}\biggr) \,, \\
 \tilde{\varphi}(e_t) &=  \biggl(1 + 1.893242666\,e_t^2 - 2.708117333\,e_t^4 
 + 0.6192474531\,e_t^6 + 0.05008474620\,e_t^8
 \nonumber \\&
 - 0.01059040781\,e_t^{10}\biggr) 
 \biggl / 
 \biggl(1 - 4.638007334\,e_t^2 + 8.716680569\,e_t^4 
 \nonumber \\&
 - 8.451197591\,e_t^6 + 4.435922348\,e_t^8
 - 1.199023304\,e_t^{10} 
 + 0.1398678608\,e_t^{12}
 \nonumber \\&
 - 0.004254544193\,e_t^{14}\biggr)\,.
\end{eqnarray}

We have verified that the above equations and expressions are consistent with Equations (3.12), (3.14), (3.15) and (3.16) of \cite{THG}. 
To compute the BBH's  secular orbital phase evolution, we solve numerically the above fully 2PN-accurate 
differential equations for $n$ and $e_t$, and the resulting temporal evolutions are imposed 
on the analytic expression for $\phi - \phi_0 = ( 1+ k) \, l$ while 
using Equation~(\ref{eqn:k_3PNl}) for $k$.
This procedure is repeated to obtain a general relativistic bound on $\gamma$.

\end{document}